\documentclass[12pt]{article}

\usepackage{multirow}
\usepackage{amsmath,amssymb,graphicx} %use drftcite in draftmode
\usepackage{pstricks}
\usepackage{bbm}
\usepackage{cite}
\usepackage{latexsym}
\newcommand{\drawsquare}[2]{\hbox{%
\rule{#2pt}{#1pt}\hskip-#2pt%  left vertical
\rule{#1pt}{#2pt}\hskip-#1pt%  lower horizontal
\rule[#1pt]{#1pt}{#2pt}}\rule[#1pt]{#2pt}{#2pt}\hskip-#2pt%  upper horizontal
\rule{#2pt}{#1pt}}% right vertical
\newcommand{\Yfund}{\drawsquare{7}{0.6}}%  fundamental
\newcommand{\centeron}[2]{{\setbox0=\hbox{#1}\setbox1=\hbox{#2}\ifdim
\wd1>\wd0\kern.5\wd1\kern-.5\wd0\fi \copy0
\kern-.5\wd0\kern-.5\wd1\copy1\ifdim\wd0>\wd1
                                   \kern.5\wd0\kern-.5\wd1\fi}}
\newcommand{\ltap}{\>\centeron{\raise.35ex\hbox{$<$}}
                           {\lower.65ex\hbox{$\sim$}}\>}
\newcommand{\gtap}{\>\centeron{\raise.35ex\hbox{$>$}}
                           {\lower.65ex\hbox{$\sim$}}\>}
\newcommand{\gsim}{\mathrel{\gtap}}
\newcommand{\lsim}{\mathrel{\ltap}}

\newcommand\ZZ{\hbox{\zfont Z\kern-.4emZ}}
\font\zfont = cmss10 %scaled \magstep1

\newcommand{\fref}[1]{fig.\ \ref{f.#1}}
\newcommand{\eref}[1]{eq.\ (\ref{e.#1})}

\newcommand{\sref}[1]{Section \ref{s.#1}}
\newcommand{\ssref}[1]{Section \ref{ss.#1}}

\newcommand{\cref}[1]{Chapter \ref{c.#1}}

\newcommand{\ba}{\begin{array}}
\newcommand{\ea}{\end{array}}

\newcommand{\beq}{\begin{eqnarray}}% can be used as {equation} or  {eqnarray}
\newcommand{\eeq}{\end{eqnarray}}
\newcommand{\beqs}{\begin{eqnarray*}}
\newcommand{\eeqs}{\end{eqnarray*}}

\newcommand{\bal}{\begin{align}} %this is glorious: arbitrary number of columns, alignment
\newcommand{\eal}{\end{align}}

\def\bi{\begin{itemize}}
\def\ei{\end{itemize}}
\def\ben{\begin{enumerate}}
\def\een{\end{enumerate}}
\def\bc{\begin{center}}
\def\ec{\end{center}}
\def\bt{\begin{table}}
\def\et{\end{table}}
\def\btb{\begin{tabular}}
\def\etb{\end{tabular}}

\def\gev{\, {\rm GeV}}
\def\kev{\, {\rm keV}}

\def\tev{\, {\rm TeV}}

\def\mass2{mass${}^2$}

\newcommand{\tr}{\mathrm T \mathrm r}

\newenvironment{tenumerate}{
\begin{enumerate}
  \setlength{\itemsep}{1pt}
  \setlength{\parskip}{0pt}
  \setlength{\parsep}{0pt}
  \vspace{-2mm}
}{\vspace{-2mm} \end{enumerate}}

\newcommand{\mprime}{{m^\prime}}

\begin{document}
\bibliographystyle{unsrt}
\begin{titlepage}
%\begin{flushright}
%{\tt hep-ph/yymmnn}
%\end{flushright}

\vskip1.5cm
\begin{center}
{\huge \bf Singlet-Stabilized Minimal Gauge Mediation}
\end{center}
\vskip0.2cm

\begin{center}
{\bf David Curtin$^1$ and Yuhsin Tsai$^{1,2}$}

\end{center}
\vskip 8pt

\begin{center}
{\it $^1$Institute for High Energy Phenomenology\\
Newman Laboratory of Elementary Particle Physics\\
Cornell University, Ithaca, NY 14853, USA } \\
\vspace*{0.3cm}
{\it $^2$Theoretical Physics Department\\
 Fermi National Accelerator Laboratory\\
  Batavia, IL 60510, USA}
\vspace*{0.3cm}

\vspace*{0.1cm}

{\tt drc39@cornell.edu, yt237@cornell.edu}
\end{center}

\vglue 0.3truecm

\begin{abstract}
We propose Singlet Stabilized Minimal Gauge Mediation as a simple ISS-based model of Direct Gauge Mediation which avoids both light gauginos and Landau poles. The hidden sector is a massive s-confining SQCD that is distinguished by a minimal $SU(5)$ flavor group. The uplifted vacuum is stabilized by coupling the meson to an additional singlet sector with its own $U(1)$ gauge symmetry via non-renormalizable interactions suppressed by a higher scale $\Lambda_{UV}$ in the electric theory. This generates a nonzero VEV for the singlet meson via the inverted hierarchy mechanism, but requires tuning to a precision $\sim (\Lambda/\Lambda_{UV})^2$, which is $\sim 10^{-4}$. In the course of this analysis we also outline some simple model-building rules for stabilizing uplifted ISS models, which lead us to conclude that meson deformations are required (or at least heavily favored) to stabilize the adjoint component of the magnetic meson.

\end{abstract}

\end{titlepage}

%%%%%%%%%%%%%%%%%%%%%%%%%%%%%%%%%%%%%%%%%%%%%%%%%%%%%%%%%%%%%%%%%%%%%%
%%%%%%%%%%%%%%%%%%%%%%%%%%%%%%%%%%%%%%%%%%%%%%%%%%%%%%%%%%%%%%%%%%%%%%%
\section{Introduction}
\label{s.intro} \setcounter{equation}{0} \setcounter{footnote}{0}
%%%%%%%%%%%%%%%%%%%%%%%%%%%%%%%%%%%%%%%%%%%%%%%%%%%%%%%%%%%%%%%%%%%%%%
%%%%%%%%%%%%%%%%%%%%%%%%%%%%%%%%%%%%%%%%%%%%%%%%%%%%%%%%%%%%%%%%%%%%%

Supersymmetry (SUSY) is an extremely elegant proposed solution to the hierarchy problem in the Standard Model (SM). However, the question of how SUSY is broken and how this breaking is communicated to the Supersymmetric Standard Model (SSM) is far from settled. Over the years many approaches have been proposed, and one of the most promising avenues is Gauge Mediation \cite{earlyGMSB, GMSB}. It automatically solves the SUSY flavor problem, since soft terms are generated by flavor-blind SM gauge interactions, and has the additional advantage of being calculable in many cases. The simplest GM models feature a single set of messengers that are charged under the SM gauge groups and couple to a SUSY-breaking hidden sector, generating the SSM soft masses through loop interactions (see \cite{GMSBreview} for a review). Many generalizations of this minimal theme exist in the literature (see, for example, \cite{earlyGMSB, GMSB, DGMother, DGMFirstNoticedLightGaugino, DGMAlsoNoticedLightGaugino, sDGMFirstNoticedLightGaugino, sDGMAlsoNoticedLightGaugino, ISS, GGM, EOGM}). For reasons of simplicity, models of Direct Gauge Mediation  are particularly appealing since they do not require a separate messenger sector; the SUSY-breaking sector talks directly with the SSM \cite{DGMother, DGMFirstNoticedLightGaugino}. By defining General Gauge Mediation as any SUSY-breaking model where the soft masses vanish as the SM gauge couplings are taken to zero, it is possible to parametrize the effects of Gauge Mediation in a very model-independent fashion \cite{GGM}.

Gauge mediation does not answer the question of how SUSY is broken, and a large variety of SUSY-breaking models can act as its hidden sector.  The most desirable scenario is a hidden sector which breaks supersymmetry dynamically.

Constructing models of dynamical SUSY breaking is extremely difficult, since the absence of any supersymmetric vacua imposes strong constraints on the theory \cite{WittenIndex}. Those requirements can be relaxed if we allow for the possibility that our universe lives in a long-lived \emph{meta-stable} SUSY-breaking vacuum, and Intriligator, Seiberg and Shih (ISS) generated enormous interest in 2006 when they demonstrated that such scenarios are fairly generic by showing that simple SUSY QCD with light quark masses can have metastable SUSY-breaking vacua near the origin of field space \cite{ISS}. In the strict sense we speak of Dynamical SUSY Breaking as scenarios where the small SUSY-breaking scale is generated dynamically, which is not the case for ISS because the small electric quark mass has to be inserted by hand. However, it does break SUSY non-perturbatively from the point of view of the UV theory and is under full calculational control using the Seiberg Duality \cite{SeibergDuality}, which together with its sheer simplicity makes it an extremely attractive model-building arena for exploring SUSY-breaking and Direct Gauge Mediation, and several attempts were made to incorporate it into phenomenologically realistic models \cite{DGMAlsoNoticedLightGaugino,KOO, ISSmesondeformNoHierarchy,ISSbaryondeform,ISSCsaba}.

The meta-stable ground state of the unmodified ISS model has an unbroken (approximate) R-symmetry that forbids gaugino masses. Breaking that symmetry spontaneously generates gaugino masses that are at least a factor of $\sim 10$ lighter than the sfermion masses. This is actually a generic feature of many Direct Gauge Mediation models, and the resulting split-SUSY-type spectrum is phenomenologically very undesirable since it exacerbates the little hierarchy problem. Explicit breaking \cite{KOO,ISSmesondeformNoHierarchy} can generate larger masses but creates new SUSY vacua and often creates a tension between reasonably large gaugino masses and stability of the ISS vacuum.

Recent work by Komargodski and Shih \cite{KStheorem} sheds light on the issue. It was shown that the leading-order gaugino mass vanishes if the SUSY-breaking vacuum is stable within the renormalizable theory. This applies to unmodified ISS, where in the magnetic theory the SUSY-vacua only show up far out in field space through non-perturbative effects. The first example of a sufficiently destabilized ISS model was \cite{KOO}, and an existence-proof of an `uplifted' model that is stabilized on a higher branch of the pseudomoduli space of massive SQCD was presented in \cite{GKK}, with later variations by \cite{GKKwSUGRA,GKKwSO10,GKKcosmo,upliftedISSbaryondeform}.

This brings us to the motivation for this paper. As is evident from the above discussion, there exists a large variety of ISS-based models of direct gauge mediation, uplifted or not. However, most of them share several shortcomings:
\begin{tenumerate}
\item Landau pole in the SM gauge couplings below the GUT-scale due to (sometimes a very large amount of) excess matter in the hidden sector.
\item The addition of nongeneric or seemingly contrived couplings and deformations, which often break global symmetries. Often there is also an unexplained partial breaking of the hidden sector flavor symmetry, both to stabilize the vacuum and to embed the SM gauge group.
\item Often severe fine-tuning to stabilize the vacuum.
\end{tenumerate}
Putting aside the fine-tuning problem for the moment, we would like to address the first two issues.  We construct a Direct Gauge Mediation model with an absolutely minimal SQCD sector which has no Landau Pole, no flavor symmetry breaking and (depending on one's judgement) no contrived deformations/couplings. The price we pay for this simplicity is the addition of the singlet sector proposed by \cite{singletsector}.  We call this model \emph{ Singlet-Stabilized Minimal Gauge Mediation}. Our UV theory will be $SU(4)_C \times SU(5)_F$ s-confining SQCD \cite{ISSNf=Nc+1} with a single quark mass scale. The IR theory has trivial gauge group and the standard model gauge group is identified with the $SU(5)_F$. There are two pseudomoduli spaces, the ISS branch with an $SU(4)$ flavor symmetry and a single uplifted branch with unbroken $SU(5)$. The vacuum is stabilized on the uplifted branch by the singlet sector. The spectrum of soft masses is precisely that of Minimal Gauge Mediation, the best possible solution from the point of view of the gaugino mass problem.

We also address an issue that may have not been explicitly discussed in the past: stabilizing an uplifted branch of massive SQCD requires \emph{two} stabilization mechanisms: one each for the adjoint and singlet components of the meson. This makes it extremely hard to avoid some meson deformations.

This paper is laid out somewhat hierarchically. In \sref{SSMGM} we outline the construction of our model and summarize all of the important results. Each summary refers to one of the later sections for details, but the essence of our work is contained in this short overview. The later chapters are organized as follows. A self-contained review of the ISS framework and related model building development is given in \sref{review}. Based on the need for two stabilization mechanisms we derive some guidelines for building uplifted ISS models in \sref{adjointinstability}. We then move on to slightly more detailed discussions of the overall vacuum structure and spectrum (\sref{vacuumstructure}), implementation of Direct Gauge Mediation to get ISS-based model of Minimal Gauge Mediation (\sref{DGM}) and the mechanism of stabilizing the uplifted vacuum (\sref{vacuumstabilization}). We conclude with \sref{conclusions}.

%%%%%%%%%%%%%%%%%%%%%%%%%%%%%%%%%%%%%%%%%%%%%%%%%%%%%%%%%%%%%%%%%%%%%%
%%%%%%%%%%%%%%%%%%%%%%%%%%%%%%%%%%%%%%%%%%%%%%%%%%%%%%%%%%%%%%%%%%%%%%%
\section{Overview of the SSMGM Model}
\label{s.SSMGM} \setcounter{equation}{0} \setcounter{footnote}{0}
%%%%%%%%%%%%%%%%%%%%%%%%%%%%%%%%%%%%%%%%%%%%%%%%%%%%%%%%%%%%%%%%%%%%%%
%%%%%%%%%%%%%%%%%%%%%%%%%%%%%%%%%%%%%%%%%%%%%%%%%%%%%%%%%%%%%%%%%%%%%

We would like to build a model of direct gauge mediation based on the ISS model \cite{ISS} that avoids both light gauginos and Landau Poles. \emph{A review of the ISS framework for metastable SUSY braking and direct gauge mediation can be found in section \sref{review}.} In this section we summarize the highlights of our model and its main physical consequences, while the details of the analysis are deferred to Sections \ref{s.adjointinstability} - \ref{s.vacuumstabilization}.

In this paper, we construct the smallest possible ISS model stabilized on the highest possible pseudomoduli space to ensure that all messengers contribute to the gaugino mass (i.e. we get Minimal Gauge Mediation). This model has no Landau Pole due to minimal excess matter and no flavor breaking. The uplifted vacuum is stabilized via a separate singlet sector, so we call this setup \emph{Singlet-Stabilized Minimal Gauge Mediation} (SSMGM).

\subsubsection*{Constructing the Magnetic Theory}
We want a trivial low-energy gauge group and an $SU(N_f) = SU(5)$ flavor symmetry. This means the electric theory must be s-confining \cite{ISSNf=Nc+1}, and strictly speaking it is inaccurate to speak of a \emph{magnetic} theory -- at low energies we use a \emph{confined} description, where the fundamental degrees of freedom are just the baryons and mesons of the original theory. However, s-confining SQCD displays similar metastable SUSY-breaking behavior as free magnetic SQCD, so in the interest of using familiar ISS-terminology we shall refer to the confined description as `magnetic' and the baryons as `magnetic squarks'.

For this choice of electric theory, pseudomoduli space of the magnetic theory only has two branches: the ISS vacuum corresponding to $k = 1$ (i.e. the magnetic squarks get a VEV) and an uplifted branch corresponding to $k = 0$ (i.e. no squarks get a VEV). If we could stabilize the uplifted branch we can identify the SM gauge group with the unbroken $SU(5)$ flavor group. The squarks would then act as a pair of Minimal Gauge Mediation messengers and generate gaugino masses at leading order in SUSY-breaking. The authors of \cite{GKK} have shown that meson deformations alone cannot achieve this stabilizations for such a small flavor group. Therefore, the price we pay for the pleasing minimality in the SQCD sector is the addition of a singlet sector with its own $U(1)$ gauge group, which spontaneously breaks the $U(1)_R$ symmetry by the inverted hierarchy mechanism \cite{InvertedHierarchy} and stabilizes the uplifted vacuum.

In the magnetic description of the ISS model, the field content is
\begin{equation}
\begin{array}{l}
\\
\mathrm{SQCD\ sector} \ \left.
\begin{array}{l}
\\ \\ \\
\end{array}\right.\\
\mathrm{singlet\ sector} \ \left.
\begin{array}{l}
\\ \\ \\ \\
\end{array}
\right.
\end{array}
\begin{array}{c|c|c|c|c}
& SU(N_f) & U(1)_B & U(1)_R & U(1)_S\\
\hline \phi^i & \Yfund & 1 & 0&0\\
\bar \phi_j & \overline \Yfund & -1 &  0&0\\
M & \mathrm{Adj} + 1 & 0 & 2&0 \\
\hline S & 1 & 0 & 0 & 1\\
\overline S & 1 & 0 & 0 & -1\\
Z & 1 & 0 & 2 & 1\\
\overline Z & 1 & 0 & 2 & -1
\end{array}
\end{equation}
where $U(1)_S$ is the gauge group of the singlet sector with coupling $g$. The complete superpotential is
\begin{equation}
\label{minISSsuperpot}
W = h \bar \phi_i M^i_j \phi^j + (- h f^2 + d S \bar S) \mathrm{Tr}M + \mprime (Z \bar S + S \bar Z)- a \frac{\det M}{|\Lambda|^{N_f - 3}}  + m_{adj} \mathrm{Tr} ({M^\prime}^2) ,
\end{equation}
where $a, h$ are unknown positive $O(1)$ numbers and $f, \mprime$ are mass scales (which can be complex) much smaller than $\Lambda$. The instanton term breaks the approximate $U(1)_R$ symmetry and restores SUSY for large meson VEVs. To explain the last term, decompose the meson into singlet and adjoint components $M = M_{sing} + M_{adj}$. The $M^\prime$ denotes the traceless part of the meson, meaning the deformation only gives a mass to $M_{adj}$. This is necessary because the singlet sector couples to $M_{sing}$ and stabilizes it away from the origin, but $M_{adj}$ is tachyonic at the origin in the uplifted pseudomoduli space. Therefore, unfortunately, we must give it a mass by hand -- this is a general feature of uplifted ISS models. \emph{For the derivation of this model-building requirement, please refer to \sref{adjointinstability}}.

\subsubsection*{The Corresponding Electric Theory \& Scales of the Model}

The electric description is an augmented massive s-confining SQCD with gauge group $SU(N_f - 1) = SU(4)$ and superpotential
\begin{equation}
\label{e.electricW}
W = \left(\tilde f + \frac{\tilde d}{\Lambda_{UV}}S \bar S \right) \mathrm{Tr}Q \bar Q + \mprime (Z \bar S + S \bar Z) + \frac{\tilde a}{\Lambda_{UV}} \mathrm{Tr}{\left(Q \bar Q\right)^\prime}^2,
\end{equation}
where $\tilde a$ is assumed to be some $O(1)$ number.
We make no attempt at explaining the origin of the small quark mass term (see \cite{ISSCsaba} for example). $\Lambda_{UV} > \Lambda$ is the scale of some UV-physics which generates the non-renormalizable $S S Q \overline Q$, $Q \overline Q Q \overline Q$ terms. The natural sizes of the IR parameters are therefore
\begin{equation}
\label{e.naturalparamsize}
d \sim \frac{\Lambda}{\Lambda_{UV}}, \ \ \ \ h \sim 1 \ \ \ \ m_{adj} \sim \frac{\Lambda^2}{\Lambda_{UV}} \sim d \Lambda.
\end{equation}

To protect the Seiberg Duality transition from the physics at scale $\Lambda_{UV}$, we conservatively require  $\Lambda_{UV} \gsim 100 \Lambda$. The masses $f$ and $m^\prime$ are free parameters as long as they are both smaller than $\sim \Lambda/100$.

A natural choice for $\Lambda_{UV}$ would be either the GUT-scale or the Planck-scale, with $\Lambda$ at least two orders of magnitude below that. In \ssref{conditionsforlocalmin} we show that decreasing $\Lambda$ much below $\sim \Lambda_{UV}/100$ makes it increasingly harder to construct uplifted metastable vacua. One can understand this quite simply as the coupling between the singlet sector and the SQCD sector becoming too weak to stabilize the magnetic meson against the effect of the instanton term, which wants to push the meson towards a supersymmetric vacuum far out in field space. This favors making $\Lambda$ as large as possible and justifies choosing two plausible scenarios for us to consider:
\begin{center}
\begin{tabular}{lcc}
& $\Lambda$ & $\Lambda_{UV}$\\
Scenario 1 & $10^{16}$ & $10^{18}$\\
Scenario 2 & $10^{14}$ & $10^{16}$
\end{tabular}
\end{center}
 (all masses in GeV), setting  $d \sim 0.01$.

\subsubsection*{The Uplifted Vacuum}

Ignoring the instanton term near the origin, $F_M$ is given by
\begin{equation}
\label{e.mesonFterm}
-F_{M^i_j}^* = h \bar \phi_i \phi^j - (h f^2 - d S \bar S)\delta^j_i.
\end{equation}
Since the first term has maximal rank $1$ and the second term has maximal rank $5$, some $F$-terms must be nonzero, breaking SUSY by the rank condition. We want to live in the uplifted vacuum, so we set $\langle \overline \phi \phi \rangle = 0$. The singlets then obtain nonzero VEV whenever $r = \sqrt{N_f h d} \ f/\mprime > 1$, in which case $F_Z, F_{\overline Z} \neq 0$ so the singlets participate in the SUSY-breaking. Some of the $\phi, \overline \phi$ are tachyonic for
\begin{equation}
\langle | M_{sing}|\rangle < \frac{\mprime}{\sqrt{h d}},
\end{equation}
but 1-loop corrections from the messengers and the singlet sector give the meson a VEV at
\begin{equation}
\langle |M_{sing}| \rangle \sim \sqrt{\frac{h}{d}} f,
\end{equation}
which is large enough to stabilize the messengers and give a viable uplifted vacuum. \emph{A complete discussion of the vacuum structure and spectrum is given in \sref{vacuumstructure}.}

\subsubsection*{Implementing Direct Gauge Mediation}

If we identify the $SU(5)$ flavor group with the SM GUT gauge group and live in the uplifted vacuum, we obtain a model of direct gauge mediation with a single pair of $(5 + \overline 5)$ messengers $\phi, \overline \phi$. Since the messengers are tachyonic for small VEVs of the meson $M$ they generate gaugino masses at lowest order in SUSY-breaking -- in fact, this is just an uplifted-ISS implementation of standard Minimal Gauge Mediation. There is no Landau pole, and the singlet degrees of freedom are all heavier than the messengers (except for the pseudomodulus, goldstino and R-axion). \emph{See \sref{DGM} for details.}

\subsubsection*{Stabilizing the Uplifted Vacuum}
The one-loop potential from the messengers tries to push the pseudomodulus (and hence the meson) towards the origin where the messengers are tachyonic, while the singlet sector contribution pushes it away from the origin. To cancel these competing contributions and create a local minimum it is necessary to adjust the ratio $\mprime/f$ to a precision of roughly
\begin{equation}
\Delta \sim \left(\frac{\Lambda}{\Lambda_{UV}}\right)^2,
\end{equation}
which is $ \sim 10^{-4}$ in our two scenarios. The tuning could be significantly reduced if one were less conservative about the separation of the two scales $\Lambda, \Lambda_{UV}$.

In our scenarios the smallness of $d$ compared to the other couplings raises the question of whether a one-loop analysis can be trusted. We show that two-loop corrections involving the larger couplings do not invalidate our analysis, because they neither influence the non-trivial part of the effective potential which generates the minimum, nor make it impossible to cancel the other smooth contributions to high enough precision so that this interesting part survives. Therefore, the meson can always be stabilized away from the origin.

Finally one must check that decays of the uplifted vacuum to both the ISS and the SUSY vacuum are suppressed enough to make the lifetime longer than the age of the universe. This is indeed the case for our model, since the bounce actions for decay to the ISS and SUSY vacua are enhanced by $(\Lambda_{UV}/\Lambda)^2$ and $\sqrt{\Lambda/f}$ respectively.

\emph{See \sref{vacuumstabilization} for a detailed discussion on stabilization of the uplifted vacuum, the effect of two-loop corrections and calculation of the vacuum lifetime.}

%%%%%%%%%%%%%%%%%%%%%%%%%%%%%%%%%%%%%%%%%%%%%%%%%%%%%%%%%%%%%%%%%%%%%%
%%%%%%%%%%%%%%%%%%%%%%%%%%%%%%%%%%%%%%%%%%%%%%%%%%%%%%%%%%%%%%%%%%%%%%%
\section{Reviewing the ISS Framework}
\label{s.review} \setcounter{equation}{0} \setcounter{footnote}{0}
%%%%%%%%%%%%%%%%%%%%%%%%%%%%%%%%%%%%%%%%%%%%%%%%%%%%%%%%%%%%%%%%%%%%%%
%%%%%%%%%%%%%%%%%%%%%%%%%%%%%%%%%%%%%%%%%%%%%%%%%%%%%%%%%%%%%%%%%%%%%
This section provides a brief summary of the ISS framework and related model building developments which form the basis of this paper. After outlining the general need for metastable SUSY-breaking in gauge mediation we review the original ISS model as well as its more recent uplifted incarnations.

\subsection{The necessity of metastable SUSY-breaking}
\label{ss.metastable}

The reasons for pursuing theories of meta-stable SUSY-breaking go beyond the significant model-building simplifications they potentially afford.

One possible argument goes as follows:
A generic theory that breaks SUSY in its ground state must have an R-symmetry (see e.g. \cite{SUSYbreakingreview} for a review). Since this forbids gaugino masses the R-symmetry must be broken. If the $R$-symmetry is only spontaneously broken one might think that the massless $R$-axion causes cosmological and astrophysical problems, necessitating explicit $R$-breaking. By the Nelson-Seiberg theorem \cite{NelsonSeibergTheorem}, this causes supersymmetric vacua to come in from infinity, making the SUSY-breaking vacuum metastable. However, \cite{Raxionmass} show that supergravity effects give the $R$-axion a mass, provided that the cosmological constant is tuned away, even if $R$-symmetry is merely spontaneously broken in the global SUSY theory. Therefore, avoiding a massless $R$-axion is \emph{not} a reason for metastable SUSY-breaking. (It is still possible that the $R$-breaking effects of gravity do in fact destabilize the SUSY-breaking vacuum, but it is not known whether the Nelson-Seiberg theorem applies in this case.)\footnote{We thank Zohar Komargodski and Jesse Thaler for pointing this out to us.}

Within the framework of Direct Gauge Mediation there is, however, another very good reason for believing in meta-stable SUSY-breaking.
%The second argument is more particular to gauge mediation but is also connected to gaugino masses.
As first noticed in \cite{DGMFirstNoticedLightGaugino}, many models of Direct Gauge Mediation suffer from very small gaugino masses compared to the sfermions. This resuls in a split-SUSY-type spectrum which reintroduces fine tuning into the Higgs Sector. Komargodski and Shih \cite{KStheorem} explored this issue in a relatively model-independent way by examining generalized O'Raifeartaigh models (renormalizable Wess-Zumino models which break supersymmetry and have canonical Kahler potentials)\footnote{\cite{KStheoremSDGM} and \cite{noncanonicalKS} extend this discussion to semi-Direct Gauge Mediation and models with non-canonical Kahler terms, respectively.}. These theories form the low-energy effective description for the hidden sector of many direct gauge mediation scenarios.

Any generalized O'Raifeartaigh model features tree-level flat directions called pseudomoduli emanating from the SUSY-breaking vacuum. The pseudomodulus is the superpartner of the Goldstino, and is stabilized somewhere on the pseudomoduli space by quantum corrections. One can always write the model in the form
\begin{equation}
W = f X + (\lambda X + m)_{ij} \psi^i \psi^j + O(\psi^3)
\end{equation}
where the scalar part of $X$ is the pseudomodulus. If we take the $\psi$'s to come in $5 + \bar 5$ pairs of $SU(5)$ then this is an example of Extra-Ordinary Gauge Mediation \cite{EOGM}. To leading order in the SUSY-breaking parameter $F/X^2$, the gaugino mass is given by
\begin{equation}
\label{e.gauginomassEOGM}
m_\lambda \propto f \frac{\partial}{\partial X} \log \det (\lambda X + m)_\mathrm{messengers}.
\end{equation}
One can show that if there are no tachyons for any choice of $X$  (i.e. the pseudomoduli space is locally stable everywhere), then $\det(\lambda X + m) = \det m$. Therefore, if the pseudomoduli space is stable everywhere, the gaugino masses vanish at leading order. Since sfermion masses are created at leading order, we have a split-SUSY spectrum.

This shows that in models of Direct Gauge Mediation, the problem of the anomalously small gaugino mass is related to the vacuum structure of the theory. In order to have a gaugino mass at leading order in SUSY-breaking, it is necessary to live in a metastable vacuum  from which lower-lying vacua (SUSY-breaking or not) are accessible within the renormalizable theory. SUSY-vacua created by non-perturbative effects far out in field space do not generate a large gaugino mass. (Notice that Minimal Gauge Mediation corresponds to $m = 0$ and a single messenger pair, so the messengers are tachyonic for $X^2 < F$ and large gaugino masses are generated.)

Since the gaugino mass formula \eref{gauginomassEOGM} is only valid to lowest order in $F/X^2$ one might think that sizeable gaugino masses could be generated for large SUSY-breaking. We conducted a small study within the framework of Extra-Ordinary Gauge Mediation using both analytical and numerical techniques, and like many before us  \cite{GMSBreview, smallgauginolargesusybreaking}, we conclude that the gaugino-to-sfermion mass ratio $m_\lambda/m_{\tilde f}$ can not be tuned to be larger than $\sim 1/10$ due to a curious numerical suppression of the subleading terms.

\subsection{The ISS Model}
\label{ss.ISS}

The authors of \cite{ISS} considered UV-free SQCD with an $SU(N_c)$ gauge group and $N_f$ flavors of electric quarks with a small mass term
\begin{equation}
\label{e.UVquarkmass}
W = m Q^i \bar Q_i
\end{equation}
where $m \ll \Lambda$, denoting $\Lambda$ as the strong coupling scale of the theory. In the free magnetic phase $N_c < N_f < \frac{3}{2} N_c$, the low-energy theory can be studied using Seiberg Duality \cite{SeibergDuality} and is simply IR-free SQCD with an $SU(N_f - N_c)$ gauge group, a gauge singlet meson $\Phi$ and $N_f$ flavors of magnetic quarks $q, \bar q$, as well as a Landau Pole at scale $\Lambda_m$.

Writing $N = N_f - N_c < \frac{1}{3} N_f$, the symmetries of the IR theory are $[SU(N)] \times SU(N_f) \times U(1)_B \times U(1)_R$ (gauged symmetries in square brackets)\footnote{We emphasize that this $U(1)_R$ symmetry is anomalous under magnetic gauge interactions, which leads to the non-perturbative restoration of supersymmetry discussed below.}.
The fields have charges $\Phi$: $(1, \mathrm{Adj} + 1)_{0,2}$, $q$: $(N, \bar N_f)_{1,0}$ and $\bar q$: $(\bar N, N_f)_{-1,0}$. The Kahler terms of the low-energy effective degrees of freedom are canonical and the superpotential is
\begin{equation}
W = h q^a_i \Phi^i_j \bar q^j_a - h \mu^2 \Phi^i_i
\end{equation}
where $a,b,\ldots$ are gauge indices and $i,j,\ldots$ are flavor indices and $\mu \sim \sqrt{\Lambda m}$.

The $\Phi$ F-terms are
\begin{equation}
-F_{\Phi^i_j}^* = h q^a_i \bar q^j_a - h \mu^2 \delta^i_j.
\end{equation}
They cannot all be zero, since the first term has rank at most $N$ and the second term has rank $N_f \geq 3 N$, so supersymmetry is broken \emph{by the rank condition}. Expanding around the vacuum, the fields can be written as
\begin{equation}
    \label{e.ISSfielddef}
    \Phi =
    \begin{array}{ccc}
    \ \ \ \ \scriptstyle N & \! \! \!\!\! \scriptstyle {N_F-N} & \\
    \multicolumn{2}{l}{
    \left( \begin{array}{cc}V & Y \\ \overline Y & Z  \end{array} \right)}
    & \! \! \! \! \!
    \begin{array}{cc}\scriptstyle N\\\scriptstyle {N_F - N}\end{array}\\ \end{array} \ \ \ \ \
    q =
    \begin{array}{ccc}
    \ \ \ \ \scriptstyle N & \ \ \scriptstyle {N_F-N} & \\
    \multicolumn{2}{l}{
    \left( \begin{array}{cc}\mu + \chi_1 & \rho_1 \\  \end{array} \right)}
    & \! \! \! \! \!
    \begin{array}{cc}\scriptstyle N\end{array}\\ \end{array} \ \ \ \ \
    \overline q =
    \begin{array}{cc}
    \scriptstyle N &  \\
    \multicolumn{1}{l}{
    \left( \begin{array}{c} \mu + \overline \chi_1 \\ \overline \rho_1 \end{array} \right)}
    & \! \! \! \! \!
    \begin{array}{c}\scriptstyle N\\\scriptstyle {N_F - N}\end{array}\\ \end{array}\\
    \end{equation}
with matrix dimensions indicated. (Writing the squark fields with a subscript 1 will be useful for comparison to the uplifted ISS case.) The gauge symmetry is completely higgsed by the squark VEVs, and the surviving global symmetry is $SU(N)_{diag} \times SU(N_f - N) \times U(1)_{B^\prime} \times U(1)_R$. The spectrum divides into distinct sectors. (We take $\mu$ to be real for simplicity, and prime denotes traceless part.)
\begin{enumerate}\itemsep=-0mm
\item $V$ and $(\chi_1 + \bar \chi_1)$ get mass $\sim |h \mu|$ whereas $(\chi_1 - \bar \chi_1)^\prime$ gets eaten by the magnetic gauge supermultiplet via the superHiggs mechanism. This part of the spectrum is supersymmetric at tree-level.
\item $\mathrm{Tr}(\chi_1 - \bar \chi_1)$: the fermion is massless at tree level and the real part of the scalar is a classically flat direction (a pseudomodulus) which gets stabilized at zero. Both these fields obtain a mass at loop-level. The imaginary part of the scalar is the Goldstone boson of a broken $U(1)$ symmetry (a mixture of $U(1)_B$ and a diagonal $SU(N_f)$ generator) and is massless to all orders. This part of the spectrum can be made massive by gauging the $U(1)$ symmetry.%, which would explicitly break the $SU(N_f) \times U(1)_B$ flavor symmetry to $SU(N) \times $SU(N_f - N) \times U(1)^{B^\prime}$.
\item $Z$ is another pseudomodulus which gets stabilized at the origin and obtains a loop-suppressed mass.
\item $Y, \bar Y, \mathrm{Im}(\rho_1 + \bar \rho_1), \mathrm{Re}(\rho_1 - \bar \rho_1)$ get masses $\sim |h \mu|$. $\mathrm{Re}(\rho_1 + \bar \rho_1), \mathrm{Im}(\rho_1 - \bar \rho_1)$ are goldstone bosons of the broken flavor symmetry and massless
\end{enumerate}
In the original ISS model as it is defined above, both pseudomoduli are stabilized at the origin by quantum corrections and get a loop-suppressed mass. This leaves the $R$-symmetry unbroken and forbids gaugino masses, so for use in realistic scenarios of direct gauge mediation the ISS model must be modified somehow to break $R$-symmetry.

In the magnetic theory supersymmetry is restored non-perturbatively: for large $\Phi$ the squarks get a large mass and can be integrated out, leaving a pure SYM theory which undergoes gaugino condensation and has SUSY-vacua at
\begin{equation}
\langle q \rangle = 0, \ \ \ \ \ \  \langle \bar q \rangle = 0, \ \ \ \ \ \  \langle \Phi \rangle_\mathrm{SUSY} = \Lambda_m \left(\frac{\mu}{\Lambda_m}\right)^{2N/N_f-N} \mathbbm{1}.
\end{equation}
This makes the SUSY-breaking vacuum at the orgin meta-stable, but the smallness of the ratio $\mu/\Lambda_m$ guarantees that the false vacuum is parametrically long-lived.

We can understand this metastability in terms of the connection between $R$-symmetry and SUSY-breaking. The UV theory does not have an exact $R$-symmetry, but it emerges as an \emph{accidental} symmetry near the origin of the IR theory. That $U(1)_R$ is anomalous under gauge interactions and hence SUSY is restored by non-perturbative operators far out in field space. The 'smallness' of the explicit $R$-breaking near the origin guarantees that the SUSY-breaking vacuum is long-lived.

Since it will be of special interest to us later we should make a comment about the s-confining case of $N_f = N_c + 1$ \cite{ISSNf=Nc+1}. The magnetic gauge group is trivial, but SUSY is still restored far out in field space. This is due to the slightly modified dual superpotential, which includes what looks like an instanton term:
\begin{equation}
W = h \mathrm{Tr} q  \Phi \bar q - h \mathrm{Tr} \mu^2 \Phi + c \frac{1}{\Lambda^{Nf-3}} \det \Phi.
\end{equation}

\subsubsection*{Modifying the ISS model for Direct Gauge Mediation}
The ISS model looks like a promising framework for models of Direct Gauge Mediation. For example, one could gauge the unbroken $SU(N_f - N)$ flavor symmetry and embed the SM gauge group, which would give gauge charges to the (anti-)fundamentals $\rho_1, \bar \rho_1, Y, \bar Y$ and make them Extra-Ordinary Gauge Mediation \cite{EOGM} messengers, as well as the $\mathrm{Adjoint} + \mathrm{Singlet}$ $Z$. The main obstacle to such a construction is the unbroken $R$-symmetry in the original ISS model. (Many variations which break $U(1)_R$ spontaneously or explicitly have been proposed, and this discussion is not meant to be exhaustive.) Models with meson deformations \cite{KOO, ISSmesondeformNoHierarchy} add operators of the form $\sim \frac{1}{\Lambda_{UV}} Q \bar Q Q \bar Q$ in the UV theory which gives operators $\sim \Phi^2$ in the IR theory with suppressed coefficients. This explicitly breaks the $R$-symmetry and gives the singlet component of the meson a VEV, generating a gaugino mass. These deformations also make the (shifted) ISS-vacuum more unstable because new SUSY-vacua are introduced. This is per se desirable, since a nonzero gaugino mass at leading order in SUSY-breaking requires the existence of lower-lying vacua within the renormalizable theory, however there is a strong tension between making the gaugino mass somewhat comparable to the sfermion mass and making the vacuum too unstable. Another possibility is adding a baryon deformation to the superpotential, which in the example of \cite{ISSbaryondeform} involves adding a $\Lambda_{UV}^2$-suppressed operator in the UV theory and breaking $R$-symmetry spontaneously, generating a very small gaugino mass. A third possibility is the addition of a singlet-sector with its own $U(1)$ gauge symmetry to break $R$-symmetry spontaneously \cite{singletsector, ISSCsaba} via the Inverted Hierarchy Mechanism \cite{InvertedHierarchy}. This again gives a small gaugino mass, and the parameters have to be fine-tuned to stabilize the vacuum.

A common problem with these embeddings is the existence of a Landau Pole, primarily due to the existence of the SM-charged adjoint meson, and some of them also feature non-generic couplings or deformations with somewhat non-trivial flavor contractions.

\subsection{Uplifting the ISS Model}
\label{ss.upliftedISS}
It would be desirable to obtain a large gaugino mass in a direct gauge mediation model derived from massive SQCD (mSQCD). Adding meson deformations introduces new vacua and generates a gaugino mass at leading order, but the strong tension between stability and sizeable gaugino masses motivates the search for a different kind of metastability: finding a new stable vacuum in a higher branch of the pseudomoduli space of mSQCD (`uplifting' the vacuum). This possibility was first realized by Giveon, Katz and Komargodski \cite{GKK}, and we will sketch out their results below.

We start with the same UV theory as the standard ISS model \eref{UVquarkmass}. In the ISS vacuum, the squark VEV matrix has $\mathrm{rank} \langle q \bar q\rangle = N$. However, there are higher, unstable pseudomoduli spaces with $\mathrm{rank} \langle q \bar q \rangle = k$, with $k = 0,1, 2, \ldots N-1$. If we assume the squark VEV matrix has rank $k < N$ the surviving symmetry is $[SU(N-k)]\times SU(k)_D \times SU(N_f - k) \times U(1)_{B^\prime} \times U(1)_{B^{\prime \prime}}$. (As we will see we must assume that the meson is stabilized at a nonzero value, breaking the $U(1)_R$ symmetry.) We expand around the squark VEV and split the fields into representations of the unbroken symmetries:
\begin{equation}
\label{e.GKKfielddef}
\Phi =
\begin{array}{ccc}
\ \ \ \ \scriptstyle k & \! \! \!\!\! \scriptstyle {N_F-k} & \\
\multicolumn{2}{l}{
\left( \begin{array}{cc}V & Y \\ \overline Y & Z  \end{array} \right)}
& \! \! \! \! \!
\begin{array}{cc}\scriptstyle k\\\scriptstyle {N_F - k}\end{array}\\ \end{array} \ \ \ \ \
q =
\begin{array}{ccc}
\ \ \ \ \scriptstyle \ \ \ k & \ \scriptstyle {N_F-k} & \\
\multicolumn{2}{l}{
\left( \begin{array}{cc}\mu + \chi_1 & \rho_1 \\ \chi_2 & \rho_2  \end{array} \right)}
& \! \! \! \! \!
\begin{array}{cc}\scriptstyle k\\\scriptstyle {N - k}\end{array}\\ \end{array} \ \ \ \ \
\overline q =
\begin{array}{ccc}
\ \ \ \ \scriptstyle \ \ \ k & \ \scriptstyle {N-k} & \\
\multicolumn{2}{l}{
\left( \begin{array}{cc} \mu + \overline \chi_1& \overline \chi_2 \\ \overline \rho_1 & \overline \rho_2  \end{array} \right)}
& \! \! \! \! \!
\begin{array}{cc}\scriptstyle k\\\scriptstyle {N_F - k}\end{array}\\ \end{array}\\
\end{equation}
The spectrum can again be described in terms of a few separate sectors:
\begin{enumerate}\itemsep=0mm
\item $(\chi_2 \pm \bar \chi_2)$, $(\chi_1 - \bar \chi_1)$ get eaten by the massive gauge supermultiplets. Notice how $\mathrm{Tr}(\chi_1 - \bar \chi_1)$ is no longer massless at tree-level because the broken $U(1)$ is a mixture between a gauged diagonal generator and the $U(1)_B$.
\item $V$, $(\chi_1 + \bar \chi_1)$ get $F$-term mass $\sim |h \mu|$
\item The $Y, \rho, Z$-type fields can be analyzed separately. The $(Y, \bar Y, \rho_1, \bar \rho_1)$ fields obtain $Z$-dependent masses and contain $2 k (N_f - k)$ flavor goldstone bosons. In a scenario of Extra-Ordinary Gauge Mediation, these fields constitute messengers that are stable for all $Z$ and hence do not contribute to the gaugino mass. The $(\rho_2, \bar \rho_2)$ scalars are tachyonic for $|Z| < |\mu|$, as we would expect from living on an uplifted pseudomoduli space, but if $Z$ can be stabilized at a large-enough value they too are stable and act as messengers which \emph{do} contribute to the gaugino mass at leading order.
\end{enumerate}
The model-building quest is now to break $R$-symmetry and stabilize the $Z$ at a large enough value to ensure that all scalars are non-tachyonic. The authors of \cite{GKK} show that in a renormalizable Wess-Zumino model, no stable SUSY-breaking minimum exists for VEVs much above the highest mass scale of the theory. Hence stabilizing $Z > \mu$ is not feasible in the original model. They circumvent this problem by introducing a \emph{mass hierarchy} into the quark masses, with the first $k$ flavors having mass $\mu_1$ and the remaining $N_f - k$ flavors having a much smaller mass $\mu_2$. This means that the $\rho_2, \bar \rho_2$ fields are tachyonic for $Z < \mu_2 \ll \mu_1$, so stabilizing the meson VEV in the region $\mu_2 < Z < \mu_1$ is possible. They achieve this stabilization for large flavor groups and $k$ close to $N$ by adding finely-tuned meson deformations $\mathrm{Tr}(Z^2)$, $(\mathrm{Tr}Z)^2$. This model is a very important proof-of-principle and it does achieve sizeable gaugino masses as desired, but its drawbacks (Landau pole \& non-minimal hidden sector, imposed flavor-breaking mass hierarchies and meson deformations) motivated further research into stabilizing an uplifted ISS model.

\subsubsection*{Further Developments in Stabilizing Uplifted ISS}
There have since been other attempts at stabilizing the uplifted ISS model. \cite{GKKwSO10} examined the equivalent case for $SO(10)$-unified Direct Gauge Mediation, \cite{GKKwSUGRA} considered stabilization using SUGRA, and issues of cosmological vacuum selection were discussed in \cite{GKKcosmo}. Stabilization of an uplifted ISS model via baryon deformations was investigated in \cite{upliftedISSbaryondeform}, and while a stable vacuum can be achieved this way for much smaller flavor groups than the proof-of-principle case discussed above, that model also features many non-renormalizable operators with non-trivial flavor contractions and non-generic couplings, as well as an explicit breaking of the hidden sector flavor symmetry. It is in this context that we are motivated to construct an uplifted ISS model with a minimal hidden sector.

%%%%%%%%%%%%%%%%%%%%%%%%%%%%%%%%%%%%%%%%%%%%%%%%%%%%%%%%%%%%%%%%%%%%%%
%%%%%%%%%%%%%%%%%%%%%%%%%%%%%%%%%%%%%%%%%%%%%%%%%%%%%%%%%%%%%%%%%%%%%%%
\section{The Adjoint Instability}
\label{s.adjointinstability} \setcounter{equation}{0} \setcounter{footnote}{0}
%%%%%%%%%%%%%%%%%%%%%%%%%%%%%%%%%%%%%%%%%%%%%%%%%%%%%%%%%%%%%%%%%%%%%%
%%%%%%%%%%%%%%%%%%%%%%%%%%%%%%%%%%%%%%%%%%%%%%%%%%%%%%%%%%%%%%%%%%%%%
Before introducing our minimal uplifted ISS model in the next section we examine the general requirements for stabilizing a higher pseudomoduli space of massive SQCD (mSQCD). We emphasize a hitherto neglected point: there must actually be \emph{two} stabilization mechanisms, one for the singlet and one for the adjoint component of the $SU(N_f - k)$ meson $Z$. This in turn yields to some very general requirements on model building, which suggest that single-trace meson deformations are very hard to avoid in uplifted ISS models.

\subsection{The messenger contribution to $V_\mathrm{eff}(Z)$}
Let us examine an uplifted pseudomoduli space in the unmodified ISS model. (We will later add some structure to stabilize it.) The $SU(N_f - k)$ meson $Z$ is a pseudomodulus which is flat at tree-level. The leading contribution to its potential arises from one-loop corrections to the vacuum energy and can be computed using the Coleman-Weinberg formula
\begin{equation}
\label{e.Vcw}
V_\mathrm{CW} = \frac{1}{64 \pi^2} \mathrm{STr} M^4 \log \frac{M^2}{\Lambda_m^2}
\end{equation}
where $\Lambda_m$ is the cutoff of the magnetic theory. Since the tree-level spectrum of the magnetic gauge vector multiplet is supersymmetric it does not contribute at one-loop level, and by inspecting the superpotential it is clear that the masses of $V, (\chi_1 + \bar \chi_1)$ do not depend on $Z$ at tree-level. Therefore, we only need to consider the dependence of the $\rho, Y$-type spectrum on $Z$ to determine its 1-loop potential. The relevant part of the superpotential is
\begin{equation}
\frac{1}{h} W_Z = - \mu_2^2 Z^i_i + {\rho_2}_j Z^j_i \bar \rho_2^i + {\rho_1}_j Z^j_i \bar \rho_1^i + \mu_1 ( {\rho_1}_i \bar Y^i + Y_i \bar \rho_1^i)
\end{equation}
where $i,j$ are $SU(N_f - k)$ flavor indices and we hide the trivial color contractions. We have also implemented the flavor-breaking of \cite{GKK} for generality.

Since $V_\mathrm{CW}$ due to messengers is generated by single planar $Z$-loops, it can only depend on single-trace combinations of the form $\mathrm{Tr}[(Z Z^\dagger)^n]$. Furthermore, even if $\langle Z \rangle$ breaks the flavor symmetry, we can use broken $SU(N_f - k)$ generators to diagonalize $\langle Z \rangle$. Therefore it is justified to diagonalize $Z$ and treat the diagonal components separately. It is then easy to verify that $V_\mathrm{CW}^\mathrm{mess}$ slopes towards the region where $\rho_2, \bar \rho_2$ become tachyonic.

It is instructive to phrase this familiar argument in a slightly different way. Decompose the meson $Z$ into adjoint and singlet components:
\begin{equation}
\label{e.Zdecomposition}
Z^i_j = Z_{adj}^A {T^A}^i_j + Z_{sing} T_S
\end{equation}
where $T^A$ are the usual $SU(N_F - k)$-generators with a slightly modified canonical normalization due to the $Z$ being a complex scalar: $\mathrm{Tr}T^A T^B =  \delta^{AB}$, $T_S = \frac{1}{\sqrt{N_F - k}} \mathbbm{1}$. Our basic dynamical degrees of freedom are then the $(N_f - k)^2 -1$ complex fields $Z^A_{adj}$ and the flavor singlet complex field $Z_{sing}$.

We can do a flavor transformation and push all the VEV of the adjoint into one of the diagonal generators. Call this generator  $\tilde T_{adj}$ and the associated meson component $\tilde Z_{adj}$. Then
\begin{equation}
\label{e.zvev}
\langle Z \rangle = \langle \tilde Z_{adj} \rangle \tilde T_{adj} + \langle Z_{sing} \rangle T_S
\end{equation}
Replacing $Z \rightarrow \tilde Z_{adj} \tilde T_{adj} + Z_{sing} T_S$ in $\mathrm{Tr}[(Z Z^\dagger)^n]$ we can see that the expression is symmetric under exchange of $\tilde Z_{adj}$ and $Z_{sing}$, since the generators satisfy $\tr T_S \tilde T_{adj} = 0$ and $\tr T^2 = 1$.  \emph{The single-trace condition is therefore equivalent to saying that the adjoint and the singlet components make identical contributions to $V_\mathrm{CW}$}. Hence the behavior of $V_\mathrm{CW}^\mathrm{mess}$ is dictated by its dependence on the singlet component.

\subsection{Model Building Requirements for Stabilizing $Z$}
This reasoning shows that uplifted ISS models really need \emph{two} stabilization mechanisms: (i) $Z_{sing}$ must be stabilized at a nonzero VEV large enough to make the messengers non-tachyonic, and (ii) $Z_{adj}$ must be stabilized at zero VEV. If the effective potential is a single-trace object then both requirements are automatically satisfied. However, if only the singlet is stabilized (separately from the adjoint) then the vacuum will be unstable along the $Z_{adj}$ direction and the fields roll towards the lower-lying ISS vacuum. We call this phenomenon the \emph{Adjoint Instability}, and it has direct model building implications. Stabilizing the adjoint in an uplifted vacuum can be done in two ways.\vspace{-2mm}
\begin{enumerate}\itemsep=-1mm
\item Add an additional flavor adjoint. This would allow us to give $Z_{adj}$ a mass (either at tree-level or, more indirectly, at 1-loop).
\item Alternatively, to obtain an effective $Z_{adj}^2$ term we can do one of the following:
    \begin{enumerate}\itemsep=0mm \vspace{-2mm}
    \item Break R-symmetry explicitly by adding meson deformations like $(\tr Z)^2, \tr(Z^2)$.
    \item Break R-symmetry spontaneously, e.g. by introducing a field $A$ with R-charge $-2$ which somehow gets a VEV and gives a mass to the adjoint via the coupling $W \supset A M M$.
    \end{enumerate}
\end{enumerate}\vspace{-1mm}
Adding a flavor adjoint would greatly exacerbate the Landau Pole Problem, and Option 2 (b) is not very attractive because the corresponding operators in the UV would be even more non-renormalizable than meson deformations. (Not to mention the additional machinery required to give $A$ its VEV.) 2 (a) seems like the best solution.

This was also the path taken by the authors of \cite{GKK}. They stabilize the vacuum by effectively adding a \emph{single}-trace deformation $\mathrm{Tr}(Z^2)$. This deformation treats the singlet and the adjoint equally, and therefore stabilizing the singlet also stabilizes the adjoint. To lift the mass of $Z_{adj}$ and avoid a Landau Pole below $\Lambda_m$ without destabilizing the nonzero singlet VEV they must then add another single-trace deformation $\mathrm{Tr}(Z_{adj}^2)$. \cite{upliftedISSbaryondeform} must also include a single-trace meson deformation to stabilize the meson.

This leads us to conclude that meson deformations $\sim \frac{1}{\Lambda_{UV}} Q \bar Q Q \bar Q$ are extremely hard to avoid in mSQCD models with meta-stable SUSY-breaking vacua on uplifted pseudomoduli spaces.

%%%%%%%%%%%%%%%%%%%%%%%%%%%%%%%%%%%%%%%%%%%%%%%%%%%%%%%%%%%%%%%%%%%%%%
%%%%%%%%%%%%%%%%%%%%%%%%%%%%%%%%%%%%%%%%%%%%%%%%%%%%%%%%%%%%%%%%%%%%%%%
\section{Vacuum Structure \& Spectrum}
\label{s.vacuumstructure} \setcounter{equation}{0} \setcounter{footnote}{0}
%%%%%%%%%%%%%%%%%%%%%%%%%%%%%%%%%%%%%%%%%%%%%%%%%%%%%%%%%%%%%%%%%%%%%%
%%%%%%%%%%%%%%%%%%%%%%%%%%%%%%%%%%%%%%%%%%%%%%%%%%%%%%%%%%%%%%%%%%%%%
Near the origin of field space there are two branches of the pseudomoduli space for this model. One is the ISS vacuum, where $k = \mathrm{rank} \langle \bar \phi \phi\rangle = 1$ and the flavor symmetry is broken down to $SU(N_f - 1)$. The other is the uplifted vacuum where $k = \mathrm{rank} \langle \bar \phi \phi\rangle = 0$, i.e. no squark VEV. To solve the gaugino mass problem we must stabilize the uplifted vacuum. Before we can analyze that stabilization, we must understand the structure of the vacuum manifold at tree-level.

\subsection{The Uplifted Vacuum $(k = 0)$}
We want to live in this uplifted vacuum without squark VEVs to solve the gaugino mass problem. With the meson decomposed into singlet and adjoint components, the superpotential is
\begin{eqnarray}
\label{e.decomposedW}
\nonumber W &=& h \ \bar  \phi \cdot  M_{adj} \cdot \phi +  m_{adj} \mathrm{Tr}(M_{adj}^2)\\
\nonumber && + \left[ \frac{h \bar \phi \phi}{\sqrt{N_f}}  + \sqrt{N_f}\left(-h f^2 + d S \bar S\right)\right] M_{sing}+  \mprime (Z \bar S + S \bar Z)\\
&&  - \frac{a}{N_f^{N_f/2}} \frac{M_{sing}^{N_f}}{|\Lambda|^{N_f - 3}} + \ldots
\end{eqnarray}
where we have omitted $\Lambda$-suppressed interactions of $M_{adj}$. For simplicity, let $f$, $\mprime$ and $\Lambda$ as well as $a, h$ be real and positive throughout this analysis. For now we simply assume that the singlet sector stabilizes $M_{sing}$ at large enough VEV to make the messengers non-tachyonic, and we postpone the detailed discussion of stabilizing the uplifted vacuum to \sref{vacuumstabilization}.

\subsubsection{Tree-level VEVs near origin of field space}
Close to the origin of field space we can ignore the instanton term in determining the VEVs of the fields. For $\langle M_{ad} \rangle = 0$ and $\langle \bar \phi \phi \rangle = 0$ we then only need to analyze the second line of \eref{decomposedW} and the tree-level potential for the singlet scalar VEVs becomes
\begin{eqnarray}
\label{e.singletpotentialGKK}
\nonumber V_{tree} &\rightarrow& \frac{1}{2} g^2 \left( |S|^2 + |Z|^2 - |\bar S|^2 - |\bar Z|^2 \right)\\
\nonumber && + \left|d \sqrt{N_f}  M_{sing} S + \mprime Z \right|^2 + \left| d \sqrt{N_f}  M_{sing}  \bar S + \mprime \bar Z \right|^2\\
&& + N_f\left| d S \bar S - h f^2 \right|^2 + \left| \mprime S \right|^2 + \left| \mprime \bar S \right|^2
\end{eqnarray}
The first line is the $D$-term potential for the singlet $U(1)_S$ gauge group, and can be set to zero by imposing $|S| = |\bar S|, |Z| = |\bar Z|$. The $F_{S, \bar S}$-terms in the second line vanish for
\begin{equation}
\label{e.Zvev}
\langle Z \rangle = - d  \sqrt{N_f} \frac{\langle M_{sing} S\rangle}{\mprime}, \ \ \ \ \ \ \ \langle
\bar Z \rangle = -  d \sqrt{N_f} \frac{\langle M_{sing} \bar S\rangle}{\mprime}.
\end{equation}
This leaves the last line as the potential for $S, \bar S$, which implies
\begin{equation}
\label{e.SvevGKK}
\langle S \bar S \rangle = \frac{h f^2}{d} - \frac{\mprime^2}{d^2 N_f} \ \ \ \ \mathrm{whenever} \ \ \ \ r > 1 \ \ \ \ \mathrm{where} \ \ \  r = \sqrt{N_f h d} \frac{f}{\mprime}.
\end{equation}
(Often it is convenient to parametrize $f$ in terms of $r$, as we will see below.) We will assume that this condition is satisfied so that the singlets get a VEV and break the $U(1)_S$ gauge symmetry, which in turn can lead to spontaneous $R$-symmetry breaking via the inverted hierarchy mechanism. The only nonzero F-terms are
\begin{equation}
\langle F_{M_{sing}} \rangle = - \frac{\mprime^2}{d \sqrt{N_f}}, \ \ \ \ \ \ \ \langle F_{Z, \bar Z} \rangle =  \frac{\mprime^2}{d \sqrt{N_f}} \sqrt{\frac{h f^2 d N_f}{\mprime^2} - 1},
\end{equation}
and the total vacuum energy is
\begin{equation}
\label{e.V0GKK}
\langle V_0^{k=0} \rangle = 2 h f^2 \frac{\mprime^2}{d}  - \frac{\mprime^4}{d^2 N_f}
\end{equation}

%As we see, one combination of $M_{sing}, Z, \bar Z$ VEVs is undetermined. A mixture of these fields is the pseudomodulus, and its potential is generated at 1-loop via \eref{Vcw}. We parametrize this flat direction by setting

To be precise we decompose all the  complex scalar singlets into amplitudes and phases:
\begin{equation}
S = \sigma_S e^{i \frac{\pi_S}{\langle \sigma_S \rangle}}, \ \ \ \ \ \ Z = \sigma_Z e^{i \frac{\pi_Z}{\langle \sigma_Z \rangle}}, \ \ \ \ \
M_{sing} = \sigma_{M_{sing}} e^{i \frac{\pi_{M_{sing}}}{\langle \sigma_{M_{sing}}\rangle}}, \ \  \ \ \ \mathrm{etc.}
\end{equation}
This reveals that of the 5 phases, three are fixed at tree-level whereas the other two are the $U(1)_S$ Nambu-Goldstone boson and the $R$-axion
\begin{equation}
\label{e.piR}
\pi_R = \frac{1}{F_{tot}}\left( |F_{M_{sing}}| \pi_{M_{sing}} + |F_Z| \pi_Z + |F_{\bar Z}| \pi_{\bar Z} \right) \ \ \propto \ \  \langle \sigma_{M_{sing}} \rangle \pi_{M_{sing}} + \langle \sigma_Z \rangle \pi_Z + \langle \sigma_{\bar Z} \rangle \pi_{\bar Z}
\end{equation}
respectively. Of the 5 amplitudes, one combination
\begin{equation}
\label{e.sigmaPM}
\sigma_\mathrm{PM} = \frac{1}{F_{tot}}\left( |F_{M_{sing}}| \sigma_{M_{sing}} + |F_Z| \sigma_Z + |F_{\bar Z}| \sigma_{\bar Z} \right)
\end{equation}
is undetermined at tree-level. This is the pseudomodulus, part of the scalar superpartner of the Goldstino, and since its value affects the masses of the other particles this flat direction is lifted at 1-loop, see \eref{Vcw}.

\subsubsection{Tree-level spectrum}
The $M_{adj}$ has mass $m_{adj}$. The messenger fermion and scalar masses are
\begin{equation}
m_{\phi} = \frac{h}{\sqrt{N_f}} M_{sing} \ \ \ \ \ \ m_{\tilde \phi}^2 = m_\phi^2 \pm \frac{h}{d N_f} \mprime^2.
\end{equation}
Quantum corrections need to stabilize $M_{sing}$ in a region where the messengers are not tachyonic, hence we require
\begin{equation}
\label{e.Mvevcondition}
\langle | M_{sing} | \rangle > \frac{\mprime}{\sqrt{h d}}.
\end{equation}

We define the singlet sector to mean the superfields $S, \bar S, Z, \bar Z, M_{sing}$ and the vector superfield of the $U(1)_S$. The singlet spectrum is complicated and we discuss it in detail when analyzing the stabilization of the uplifted vacuum in \sref{vacuumstabilization}. The vector multiplet eats a chiral multiplet via the superHiggs mechanism and two (one) chiral multiplets get an $F$-term ($D$-term) mass. One multiplet is massless at tree-level: it contains the Goldstino, the pseudomodulus and the R-axion.

\subsubsection{Effect of instanton term}
Turning on the instanton term creates SUSY-vacua far out in field space. The additional terms in $F_{M_{sing}}$ are easily accounted for by replacing $h f^2 \rightarrow h \tilde f^2$ in \eref{singletpotentialGKK}, where
\begin{equation}
\label{e.hftildesq}
h \tilde f^2 = h f^2 - \frac{a}{N_f^{(N_f-1)/2}} \frac{M_{sing}^{N_f-1}}{\Lambda^{N_f-3}}.
\end{equation}
(Some of the previously undetermined phases now also get a non-zero VEV, but this does not affect the one-loop stabilization of the pseudomodulus.) As $M_{sing}$ increases $h \tilde f^2 \rightarrow 0$ and hence $S, \bar S, Z, \bar Z \rightarrow 0$. Hence
\begin{equation}
\label{e.SUSYminimum}
\langle M_{sing} \rangle_\mathrm{SUSY} \sim f \left(\frac{\Lambda}{f} \right)^{(N_f-3)/(N_f-1)}  \ \ \stackrel{=}{\scriptscriptstyle N_f \rightarrow 5}  \ \ \sqrt{f \Lambda}.
\end{equation}
The small value of $f/\Lambda$ is crucial for guaranteeing longevity of the uplifted vacuum. The effect of these $R$-breaking terms as well as the stabilization of the uplifted vacuum via quantum corrections is illustrated in \fref{vacuumstructure}.

Near the origin of field space we care about the changed behavior of the R-axion and the pseudomodulus. The explicit breaking of the $R$-symmetry gives a small mass to the $R$-axion. Note that even though the large adjoint mass represents a very large explicit $R$-breaking, since the adjoint does not get a VEV it is not part of the axion. The pseudomodulus is no longer a flat direction at tree-level, but is slightly tilted away from the origin.

\begin{figure}
\hspace{-1.2cm}
\begin{tabular}{c c c c c}
\includegraphics[width=5.5cm]{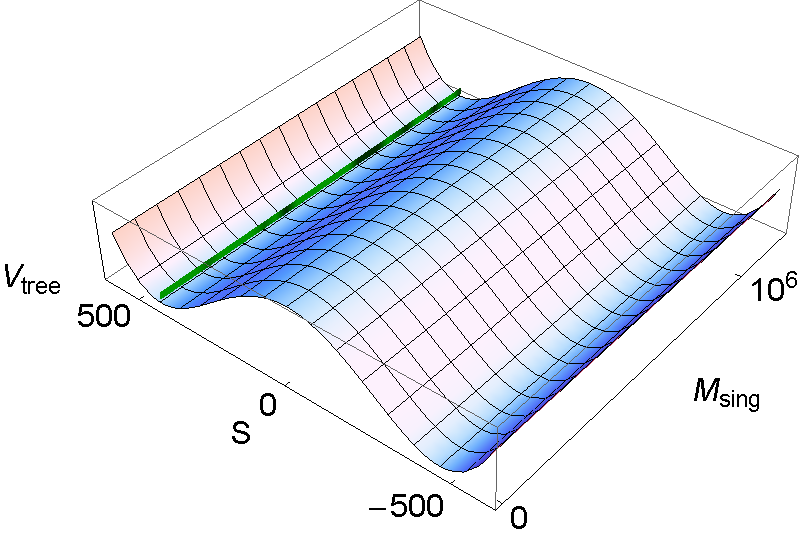}& &
\includegraphics[width=5.5cm]{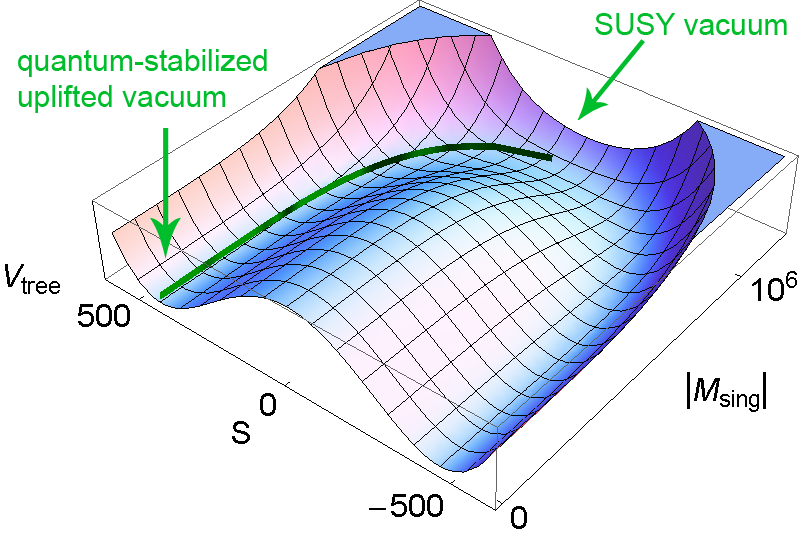}& &
\includegraphics[width=5.5cm]{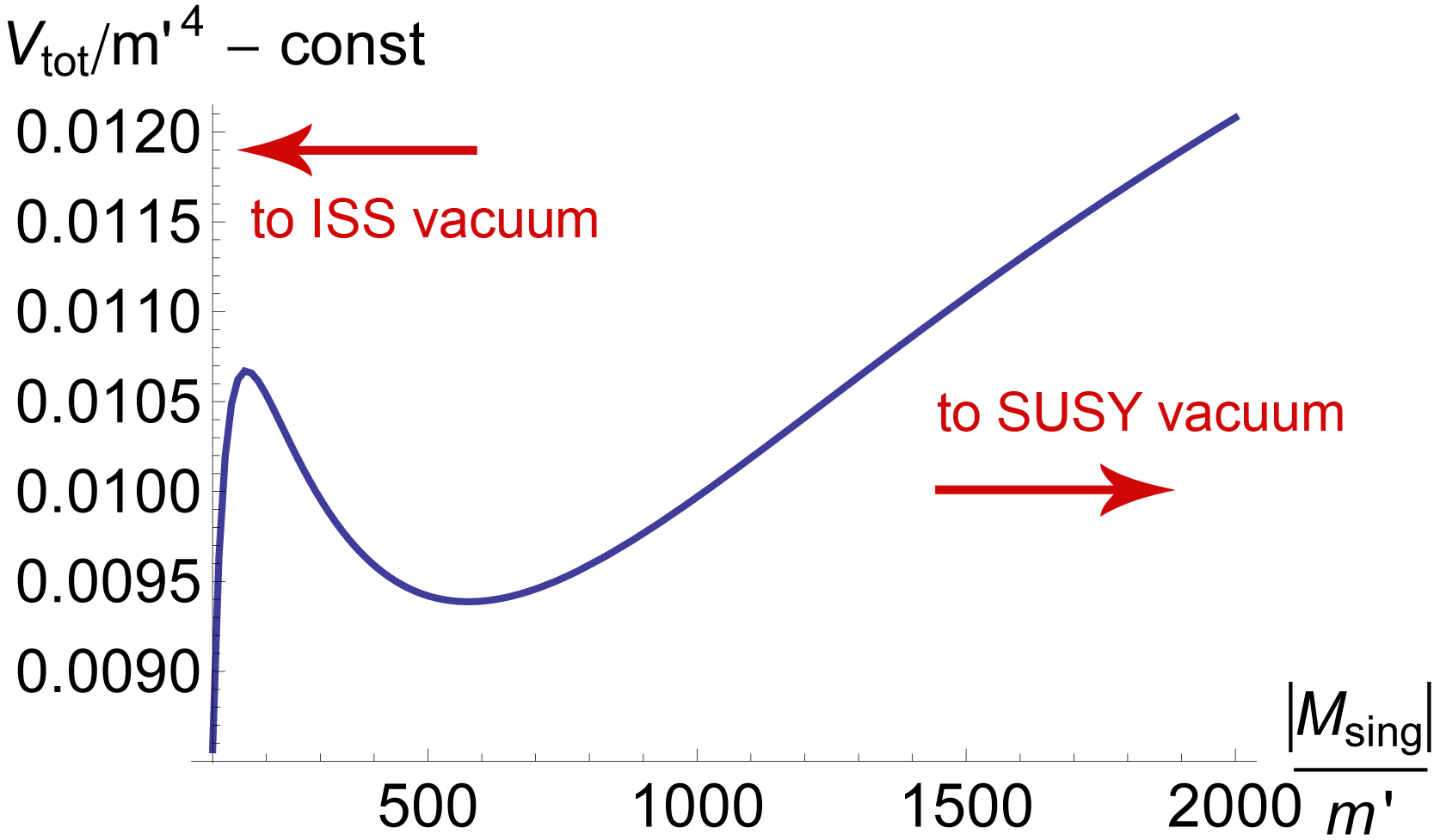}
\\(a)&&(b) && (c)
\end{tabular}
\caption{
(a) The tree-level potential without the instanton term as a function of $|M_{sing}|$ and $S$, where we have enforced tree-level VEVs $|\bar S| = |S|$ and $Z = \bar Z = -d \sqrt{N_f} M_{sing} S/\mprime$. The valley marked with a green band is perfectly flat in the $|M_{sing}|$ direction and shows that the potential has a SUSY-breaking minimum for $S^2 = \frac{h f^2}{d} - \frac{\mprime^2}{d^2 N_f}$. Note that the messengers are tachyonic for $|M_{sing}| < \mprime/\sqrt{d h}$. (b)~The same potential with the instanton term added. The minimum along the $S$-direction is approximately unchanged close to the origin but is significantly shifted as we move outwards along the $|M_{sing}|$ direction. As we walk along the the valley in the $|M_{sing}|$ direction (which now tilts slightly away from the origin) we eventually reach the SUSY-minimum at $|M_{sing}| \sim \sqrt{\Lambda f}$ and $S, Z = 0$.
(c)~We compute quantum corrections to the potential along the pseudomodulus direction, i.e. the green band in (b), by setting all fields to their VEVs in terms of $|M_{sing}|$. The vacuum is stabilized at $|M_{sing}| \sim \sqrt{h/d} \ f \ \longrightarrow\  Z, \bar Z \sim \sqrt{h/d}\ f^2/\mprime$.
The parameters used for these plots in units of $\mprime$ were $N_f =5$, $\Lambda = 3.8 \times 10^9$, $f = 63$ and $(g, d, h) = (0.02513, 0.02, 1)$.}
\label{f.vacuumstructure}
\end{figure}

%\subsubsubsection{Stabilization of the uplifted vacuum}
%As \fref{vacuumstructure}(c) demonstrates, the uplifted vacuum is stabilized by the 1-loop effects of the singlet sector.
%\begin{itemize}
%\item $M_{sing}$ is stabilized at $\sim \sqrt{d/h} f$. Therefore it is is convenient to parametrize
%    \begin{equation}
%    \label{e.Mvevparam}
%    \langle M_{sing} \rangle = b \sqrt{\frac{h}{d}} \  f, \ \ \ \ \ \mathrm{where} \ \ \ \ \ b \sim O(1).
%\item In order to have a local minimum of the total effective potential on the uplifted branch of the pseudomoduli space, the gauge goupling must be given by
%    \begin{equation}
%    g^2 = \frac{h^2 + d^2 N_f t}{4 (r^2 - 1)},
%    \end{equation}
%    where $r = \sqrt{h d N_f} f/\mprime > 1$ to have singlet VEVs, and $t$ must be in the range $\sim (0.5, 1)$. This means the gauge coupling must be tuned to a precision of
%    \begin{equation}
%    \frac{\delta g}{g} \sim \frac{d^2}{h^2}.
%    \end{equation}
%    Furthermore, $r \lsim 10^2, 10^1$ for Scenario 1 and 2 respectively.
%\item
%\end{itemize}
%This is discussed in detail in \sref{vacuumstabilization}. For now these key results are sufficient to continue our discussion of the model.

\subsubsection{Tree-level zero modes}
The fermionic component of the tree-level zero mode multiplet is the Goldstino, which is eaten by the Gravitino once SUSY is gauged and gets the familiar mass
\begin{eqnarray}
\begin{array}{cccccr}
m_{\tilde G} &=&  \displaystyle \frac{F_{tot}}{\sqrt{3} M_{pl}^*} &\approx&\displaystyle  0.4 \  \frac{r}{d} \ \frac{\mprime^2}{M_{pl}^*} + O(r^{-1}) &\ \ \ \mathrm{for}\ N_f = 5
\end{array},
\end{eqnarray}
where $M_{pl}^* = (8 \pi G_N)^{-1/2} = 2.4 \times 10^{18} \gev$ is the reduced Planck Mass. (Since $r = \sqrt{h d N_f} f/\mprime > 1$ and $d \ll 1$, it is often instructive to expand for large $r$ or large $f/\mprime$.)
The scalar components are the pseudomodulus and the R-axion (eqns \ref{e.piR}, \ref{e.sigmaPM}). To compute the 1-loop potential for the pseudoflat direction we  set all their phases to their tree-level VEV or zero
%make all the fields real (equivalent to setting all phases to their tree-level VEV or zero)
and express $\langle Z\rangle, \langle \bar Z \rangle$ in terms of $M_{sing}$, which gives $V_{\mathrm{CW}}(M_{sing})$. We emphasize that $|M_{sing}|$ is not the pure pseudomodulus, but its value parametrizes where we are along the pseudo-flat direction in field space.\footnote{To avoid clutter, we omit the absolute value signs around $M_{sing}$ from now on -- they are understood when we talk about $M_{sing}$ as parameterizing the pseudomodulus direction.} This gives $V_{eff}(M_{sing}) = V_{tree}(M_{sing}) + V_\mathrm{CW}(M_{sing})$. As per the discussion above, the first term is nonzero if we include the instanton term. Minimizing $V_{eff}$ gives $\langle M_{sing} \rangle$ and hence $\langle Z \rangle,\langle \bar Z\rangle,\langle S\rangle,\langle \bar S\rangle$. To compute the derivative $V_{eff}$ along the flat direction we differentiate with respect to $M_{sing}$ and multiply by a scaling factor $F_{M_{sing}}/F_{tot}$ to account for the fact that moving by $\delta$ along the $M_{sing}$ axis moves us by $\delta \sqrt{(F_Z/F_{M_{sing}})^2 + (F_{\bar Z}/F_{M_{sing}})^2 + 1}$ along the pseudo-flat direction. Hence we obtain the pseudomodulus mass as
\begin{equation}
m_{PM}^2 = \left( \frac{F_{M_{sing}}}{F_{tot}}\right)^2 \frac{d^2 V_{eff}}{d (M_{sing})^2}.
\end{equation}
A similar argument holds for the $R$-axion mass if we restore the undetermined phases in the tree-level potential. To ensure that we move along the correct direction in field space we impose $\pi_{Z, \bar Z} = \frac{F_Z}{F_{M_{sing}}} \pi_{M_{sing}}$, differentiate with respect to $\pi_{M_{sing}}$ and apply the same scaling factor.

These masses can be readily estimated. As we will see in \sref{vacuumstabilization}, $M_{sing}$ is stabilized at $\sim \sqrt{d/h} f$. Therefore it is is convenient to parametrize
\begin{equation}
\label{e.Mvevparam}
\langle M_{sing} \rangle = b \sqrt{\frac{h}{d}} \  f, \ \ \ \ \ \mathrm{where} \ \ \ \ \ b \sim O(1).
\end{equation}
To obtain the $R$-axion mass we differentiate the tree-level potential with all VEVs subbed in.
%This gives
%\begin{equation}
%\frac{d^2 V_{tree}}{d ( \pi_{sing})^2} = \left[ (N_f-1)(N_f-2)N_f^{-(N_f-1)/2}\right] a \ b^{N_f-3} \ \frac{\mprime^2}{d} \  \left( \frac{f}{\Lambda} \right)^{N_f-3}.
%\end{equation}
To lowest order in $1/r$ and $1/\Lambda$ we find that
\begin{equation}
\label{e.mRestimate}
\frac{F_{M_{sing}}}{F_{tot}} \approx - \frac{1}{\sqrt{2 d h N_f}} \frac{\mprime}{f} \ \ \ \ \longrightarrow \ \ \ \ \
m_R \approx 0.2 \ b \ \sqrt{\frac{a}{d^3}} \
\frac{\mprime^2}{\Lambda} \ \ \ \ \ \ \  \mathrm{for} \ N_f = 5.
\end{equation}
To estimate the mass of the pseudomodulus we pre-empt another result from \sref{vacuumstabilization}. The rough \emph{scale} of the second derivative of the 1-loop potential is
\begin{equation}
\label{e.DDVcwestimate}
\left|\frac{d^2 V_{\mathrm{CW}}}{d ( M_{sing})^2}\right| \sim \frac{1}{16 \pi^2} \frac{\mprime^4}{\langle M_{sing} \rangle^2}
\end{equation}
(where $Z, \bar Z \rightarrow Z(M_{sing}) = - d \sqrt{N_f} M_{sing} \langle S \rangle/\mprime$).  To lowest order in $1/r$ this yields
\begin{equation}
\label{e.mPMestimate}
m_{PM} \sim \frac{1}{\sqrt{32 N_f}\ \pi} \frac{\mprime}{b h} \left(\frac{\mprime}{f}\right)^2 \ \  \approx  \ \ 0.1 \ \frac{d}{b}\  \frac{\mprime}{r^2}  \ \ \ \ \ \ \  \mathrm{for} \ N_f = 5.
\end{equation}
Notice the $\mprime/f$ suppression, simply due to the fact that if $f \gg \mprime$ then $F_{M_{sing}} \ll \langle M_{sing} \rangle^2$ (similarly for $Z$, $\bar Z$) and SUSY-breaking is weak. (Effectively this can also be seen as a suppression for small $d$, since decreasing $d$ increases the minimum size of $f$ to ensure \eref{SvevGKK} is satisfied.)

\subsection{The ISS Vacuum $(k = 1)$}
Since this is very similar to a standard $(N,N_f) = (1,5)$ ISS vacuum we will use the notation of \ssref{ISS} (except for renaming the $SU(N_f - N)$ meson $Z \rightarrow \tilde M$ to avoid confusion with the singlets $\bar Z, Z$) and split up the meson according to \eref{Zdecomposition}. The squark VEV $\langle \bar \chi_1 \chi_1 \rangle = f^2 - \frac{d}{h} S \bar S$ sets $F_V = 0$, with all other SQCD-sector VEVs zero (except $\tilde M_{sing}$). This gives the same singlet potential as \eref{singletpotentialGKK} with $N_f \rightarrow N_f - 1$. Therefore the VEVs at tree-level close to the origin are $\langle|S|\rangle =\langle |\bar S|\rangle, \langle |Z| \rangle = \langle |\bar Z|\rangle$, $\langle Z\rangle = -  \sqrt{N_f - 1} \frac{\langle \tilde M_{sing} S\rangle }{\mprime}$, and
\begin{equation}
\label{e.SvevISS}
\langle S \bar S \rangle = \frac{h f^2}{d} - \frac{\mprime^2}{d^2 (N_f - 1)} \ \ \ \ \mathrm{whenever} \ \ \ \ h f^2 > \frac{\mprime^2}{(N_f - 1) d}.
\end{equation}
If this condition is not satisfied the singlets do not get a VEV and we have a standard ISS vacuum. If we assume the condition holds (slightly stronger than \eref{SvevGKK}), then $\langle \bar \chi_1 \chi_1 \rangle = \mprime^2/(d h {N_f - 1})$, meaning the scale of the squark VEV is given by $\mprime$ instead of $f$. The total vacuum energy is
\begin{equation}
\label{e.V0ISS}
\langle V_0^{k=1} \rangle = 2 h f^2 \frac{\mprime^2}{d}  - \frac{\mprime^4}{d^2 (N_f - 1)}
\end{equation}
The SQCD spectrum is the same as ISS with mass scale $ \sim \mprime$, and the singlet spectrum looks very similar to the uplifted case. We will not dwell on analyzing this vacuum, we only needed to know the potential difference
\begin{equation}
\label{e.ISSGKKenergydifference}
\Delta V_0 \equiv \langle V_0^{k=0} \rangle - \langle V_0^{k=1}  \rangle = \frac{\mprime^4}{d^2} \frac{1}{N_f (N_f - 1)}
\end{equation}
to calculate the uplifted vacuum lifetime in \ssref{lifetimecalculation}.

%%%%%%%%%%%%%%%%%%%%%%%%%%%%%%%%%%%%%%%%%%%%%%%%%%%%%%%%%%%%%%%%%%%%%%
%%%%%%%%%%%%%%%%%%%%%%%%%%%%%%%%%%%%%%%%%%%%%%%%%%%%%%%%%%%%%%%%%%%%%%%
\section{Direct Gauge Mediation}
\label{s.DGM} \setcounter{equation}{0} \setcounter{footnote}{0}
%%%%%%%%%%%%%%%%%%%%%%%%%%%%%%%%%%%%%%%%%%%%%%%%%%%%%%%%%%%%%%%%%%%%%%
%%%%%%%%%%%%%%%%%%%%%%%%%%%%%%%%%%%%%%%%%%%%%%%%%%%%%%%%%%%%%%%%%%%%%
If we weakly gauge the $SU(5)$ flavor group and identify it with the SM GUT gauge group, this model realizes Minimal Gauge Mediation with a single $5 \oplus \bar 5$ messenger pair:
\begin{equation}
W_{eff} = X \bar \phi_i \phi^i,
\end{equation}
where the SUSY-breaking spurions $X = X + \theta^2 F$ is given by
\begin{equation}
X = \frac{h}{\sqrt{N_f}} M_{sing} \  \ \ \ \rightarrow \ \ \ \ F = \frac{h}{\sqrt{N_f}} F_{M_{sing}} = - \frac{h}{d N_f} \mprime^2.
\end{equation}
Gaugino and sfermion masses are generated via the well-known 1- and 2-loop diagrams and are parametrically the same size, solving the Gaugino Mass Problem. Using equations (\ref{e.naturalparamsize}), (\ref{e.Mvevparam}) and (\ref{e.SvevGKK}) we can see that SUSY-breaking is weak:
\begin{equation}
\left| \frac{X^2}{F} \right| = \left(\frac{f}{\mprime}\right)^2 h^2 b^2 > \frac{h b^2}{d N_f} \gg 1,
\end{equation}
and therefore the soft masses are given by the usual simple expression
\begin{equation}
\label{e.msoft}
m_{soft} \sim \frac{\alpha}{4 \pi} \left| \frac{F}{X} \right|.
\end{equation}
Requiring TeV-scale soft masses sets $|F/X| \sim 100 \tev$. This determines the scale of $\mprime$ (and hence $f$):
\begin{equation}
\label{e.matchmprime}
\mprime \sim \left| \frac{F}{X} \right| b r,
\end{equation}
which sets the messenger mass at
\begin{equation}
X \sim b^2 r^2 \frac{h}{d N_f} \left| \frac{F}{X} \right| \sim r^2 \ \frac{0.01}{d} \times (10^7 \gev)
\end{equation}
in the scenarios we are considering. The pseudomodulus, and Goldstino mass scales are
\begin{eqnarray}
m_{PM} & \sim & \frac{1}{r}\ \left(\frac{d}{0.01}\right) \times (10 \gev)\\
m_{\tilde G} & \sim &  b^2 \  r^3 \ \left(\frac{0.01}{d}\right) \times (\kev).
\end{eqnarray}
The field theory contribution to the $R$-axion mass is
\begin{equation}
m_{R}  \sim b^3 \ r^2 \ \left(\frac{0.01}{d}\right)^{3/2} \ \frac{\Lambda_{GUT}}{\Lambda} \times (100 \kev).
\end{equation}
Depending on the size of $r$ and $b$ as well as the choice of scenario, this can be smaller or larger than the BPR contribution \cite{Raxionmass}.

Again using results from the next section for convenience, the mass of the singlet vector multiplet is similar to the messenger mass whereas the other singlets (with the exception of the tree-level zero modes) obtain a smaller mass $\sim r^2 |F/X|$. Stabilizing the uplifted vacuum in scenarios 1 and 2 requires $r \lsim 10^2$ and $r \lsim 10^1$ respectively, but saturating the former bound gives a very heavy gravitino and reintroduces the SUSY flavor problem. Therefore $1 < r \lsim 10^1$ is the relevant parameter range for our model.

Since the adjoint meson gets a mass that is only a few orders of magnitude below the duality transition scale $\Lambda$, which itself is either at or close to the GUT-scale, there is no Landau Pole in our model. (Scenario 2 is also an example of \emph{deflected unification} \cite{DeflectedUnification}.) However, we emphasize that due to the minimality of this hidden sector such a heavy adjoint is not required to solve the Landau Pole Problem -- if the adjoint mass was generated by some other mechanism it could be as low as $\sim 10 - 100 \tev$.

%%%%%%%%%%%%%%%%%%%%%%%%%%%%%%%%%%%%%%%%%%%%%%%%%%%%%%%%%%%%%%%%%%%%%%
%%%%%%%%%%%%%%%%%%%%%%%%%%%%%%%%%%%%%%%%%%%%%%%%%%%%%%%%%%%%%%%%%%%%%%%
\section{Stabilizing the Uplifted Vacuum}
\label{s.vacuumstabilization} \setcounter{equation}{0} \setcounter{footnote}{0}
%%%%%%%%%%%%%%%%%%%%%%%%%%%%%%%%%%%%%%%%%%%%%%%%%%%%%%%%%%%%%%%%%%%%%%
%%%%%%%%%%%%%%%%%%%%%%%%%%%%%%%%%%%%%%%%%%%%%%%%%%%%%%%%%%%%%%%%%%%%%
We now examine how the singlet sector originally proposed in \cite{singletsector} stabilizes the uplifted vacuum. The stabilization is possible due to the singlet sector's $U(1)_S$ gauge group \cite{InvertedHierarchy}, which can supply a negative coefficient to the logarithmic dependence of $V_\mathrm{CW}$ and push the minimum away from the origin beyond the region where the messengers are tachyonic. We perform this analysis to 1-loop order even though $d \ll h$ and 2-loop effects from $h$ might be competitive. This will be justified in \ssref{validityof1loop}. For simplicity we set $a = 1$ throughout.

The effective potential is given by
\begin{equation}
V_{eff} = V_{tree} + V_\mathrm{CW},
\end{equation}
where all tree-level VEVs and masses are expressed as functions of $M_{sing}$, which parametrizes the pseudomodulus VEV. $V_{tree}$ is easily obtained by combining equations (\ref{e.V0GKK}) and (\ref{e.hftildesq}).
\begin{equation}
\label{e.VtreeInstanton}
V_{tree} = \frac{2 h f^2 \mprime^2}{d} - \frac{\mprime^4}{d^2 N_f} - \frac{2 \mprime^2}{d} \frac{a}{N_f^{(N_f-1)/2}} \frac{M_{sing}^{N_f-1}}{\Lambda^{N_f-3}}.
\end{equation}
This slopes away from the origin due to the effect of the instanton term.  $V_\mathrm{CW}$ is computed by obtaining the mass spectrum \emph{without the effects of the instanton term}\footnote{If the instanton term is so large that its backreaction significantly affects the 1-loop potential, its tree-contribution will be so large as to erase any minima created by $V_\mathrm{CW}$ anyway.} and using \eref{Vcw}.

\subsection{Organizing the Spectrum \& Contributions to $V_{\mathrm{CW}}$}
All nonzero tree-level masses depend on the value of the pseudomodulus, parametrized by the value of $M_{sing}$ by imposing $Z = \bar Z = - d \sqrt{N_f} S M_{sing}/\mprime$. It is helpful to express all masses in units of $\mprime$ and define the following set of parameters:
\begin{eqnarray}
x = d \sqrt{N_f} \frac{|M_{sing}|}{\mprime}\ , \ \ \  \ \ r = \sqrt{h d N_f} \frac{f}{\mprime}\ , \ \ \ \ q = \frac{4}{N_f} \frac{g^2}{d^2} (r^2-1)\ , \ \ \  \ \ p = \frac{h}{d N_f}.
\end{eqnarray}
In this parametrization, $h$ just rescales the other variables. $r > 1$ is required for singlet VEVs. This parametrization has the advantage that the masses in every split supermultiplet depend only on $x$ and one of the $r, q, p$ parameters. This allows us to study the different $V_\mathrm{CW}$ contributions independently as functions of just two variables each.
\begin{itemize}\itemsep=1mm
\item The messenger masses can be written as $m_F^2 = p^2 x^2$ and $m_S^2 = p^2 x^2 \pm p$, and are tachyonic for $x < 1/\sqrt{p}$ (recall that we use $\mprime$ as our unit of mass in this parameterization). In the leading-log approximation for large $x$ their contribution to the 1-loop potential is $V_\mathrm{CW}^{mess} \approx \frac{1}{64 \pi^2} 8 N_f p^2 \log x$. (We will ignore additive constants to the potential.)

\item Two singlet chiral supermultiplets have $F$-term masses that depend only on $r$ and $x$. For large $x$ their masses go as $\sim x$ and $\sim 1/x$, so we denote them $R_{heavy}$ and $R_{light}$ respectively. The contribution $V_\mathrm{CW}^{Rheavy}$ stands out because it is the only one that always has a local minimum, located at  $x \approx 1.3 r - 1$ to a very good approximation.

    For most values of the parameters the other contributions to the 1-loop potential wash out this minimum and the uplifted pseudomoduli space is not stabilized. However, if the other components cancel to high enough precision then the minimum survives and is located at $\langle x \rangle \sim r > 1$ (see \eref{SvevGKK}). This justifies the parametrization
    \begin{equation}
    \label{e.Mvevparam2}
    \langle M_{sing} \rangle = b \sqrt{\frac{h}{d}} f \ \ \ \ \mathrm{where} \ \ \ \ b = O(1).
    \end{equation}

    For large $x$ the light multiplet does not contribute to $V_\mathrm{CW}$, whereas $V_\mathrm{CW}^{Rheavy} \approx \frac{1}{64 \pi^2} 4 \log x$. Near the local minimum of the total 1-loop potential, their masses to lowest order in $1/r$ are $m^2_{\stackrel{\scriptscriptstyle  Rheavy}{\scriptscriptstyle  Rlight}} \approx \frac{1}{2}\left(4+b^2 \pm b \sqrt{8 + b^2}\right) r^2$.

\item One chiral and one vector multiplet get masses from the $U(1)_S$ $D$-term, both $\sim x$ for large $x$. Call them $Q_\mathrm{vector}$ and $Q_\mathrm{chiral}$. In the leading-log approximation the contributions to the 1-loop potential are $V_\mathrm{CW}^{Qvector} \approx \frac{1}{64 \pi^2} (-8 q) \log x$ and $V_\mathrm{CW}^{Qchiral} \approx \frac{1}{64 \pi^2} 4 \log x$. Near the local minimum of $V_\mathrm{CW}$, their masses to lowest order in $1/r$ are \\$m_{Qvector}^2 \approx 4 b^2 g^2 r^4/(d^2 N_f)$ and $m_{Qchiral}^2 \approx b^2 r^2$.
\end{itemize}
Adding all the contributions together, we see that the total 1-loop potential in the leading log approximation valid for `large' field values of $M_{sing}$ corresponding to $x \gsim O(1) \gg 1/\sqrt{p}$ is
\begin{equation}
\label{e.Vcwleadinglog}
V_\mathrm{CW} \approx \frac{1}{8 \pi^2}(1 - t) \log x,
\end{equation}
where it will be convenient to define
\begin{equation}
t = q - N_f p^2.
\end{equation}

\subsection{Conditions for local minimum}
\label{ss.conditionsforlocalmin}
% t has to be between 1/2 and 1... immediately see how finely tuned g must be
% express gauge coupling in terms of epsilon g & show some plots
The leading-log approximation is excellent for $V_\mathrm{CW}^{mess}$ and $V_\mathrm{CW}^Q$, even as close to the origin as $x \sim \langle x \rangle$. Hence we can understand the tuning required for stabilizing the uplifted vacuum as follows. Imagine starting out with a choice of parameters for which there is a local minimum of $V_\mathrm{CW}$. If we then increase $t$, the coefficient of the logarithm in the potential decreases until the minimum is wiped out and the potential just slopes towards the SUSY-minimum. Conversely, if we decrease $t$ the coefficient of the logarithm increases and the minimum gets pushed towards the origin, eventually disappearing into the region where the messengers are tachyonic. Therefore having a local minimum requires $t \in (t_{min}, t_{max})$, where $t_{min, max}$ are $O(1)$ functions of the other parameters. Expressing the singlet-sector gauge coupling in terms of $t$,
\begin{equation}
g(t)^2 = \frac{h^2 + d^2 N_f t}{4(r^2-1)},
\end{equation}
translates this condition into a required tuning for $g$. However, it is more instructive to recast the stabilization requirement as a constraint on the mass ratio
\begin{equation}
\label{e.mprimetuning}
\left(\frac{\mprime}{f}\right)^2 = 4 g^2 N_f\frac{d}{h}\left(1 - \frac{d^2}{h^2} N_f t\right) + O(g^4) + O(d^5).
\end{equation}
We can see immediately that even if $t$ is allowed to take on an $O(1)$-range of values to guarantee a local minimum, $\mprime/f$ must actually be adjusted to a precision of
\begin{equation}
\Delta \sim \frac{d^2}{h^2} \sim \left(\frac{\Lambda}{\Lambda_{UV}}\right)^2.
\end{equation}
This is $\sim 10^{-4}$ in the two scenarios we are considering but could be significantly larger if one were less conservative about the separation of scales for $\Lambda, \Lambda_{UV}$. Tuning of this order of severity is typical in uplifted models that are stabilized by 1-loop corrections, and we make no attempt to explain it here. It would be very interesting to investigate whether such a mass ratio might be generated by some kind of UV-completion, but it lies beyond the scope of this paper.

What is the actual allowed range of $t$? If we switch off the instanton term then there can be no minima of $V_\mathrm{CW}$ if the coefficient of the logarithm is negative for large $x$. Hence $t^{approx}_{max} = 1$. To find the smallest allowed value of $t$ we numerically investigate the behavior of $V_\mathrm{CW}$ and we find that $t^{approx}_{min} \geq 1/2$, with the inequality becoming saturated for $r \gsim 10$.  Switching on the instanton term has the effect of reducing $t_{max}$ from the approximate value of $1$, since the $V_{tree}$ contribution has negative slope and increasing $t$ beyond $t_{min}$ causes the overall potential to have negative slope before we reach $t = 1$. This effect is more pronounced for larger $r$, since increasing $f/\Lambda$ increases the effect of the instanton term.

To understand this in more detail we studied the complete $V_{eff}$ numerically. By fixing $|F/X|$ in \eref{msoft} at $100 \tev$ one can find $t_{min}, t_{max}$ as functions of $r$ for various values of $d$ and $h$ in scenarios 1 and 2, see \fref{tplot}. As expected the instanton term does not have a significant effect on $t_{min}$ but decreases $t_{max}$ from 1 with increasing severity for larger $r$. This effectively defines a maximum value of $r$ for which there can still be a local minimum of $V_{eff}$, and  $r_{max}$ appears approximately $\propto d$ for fixed $\Lambda$, $\Lambda_{UV}$.

\begin{figure}
\begin{center}
\begin{tabular}{cc}
\includegraphics[width=8cm]{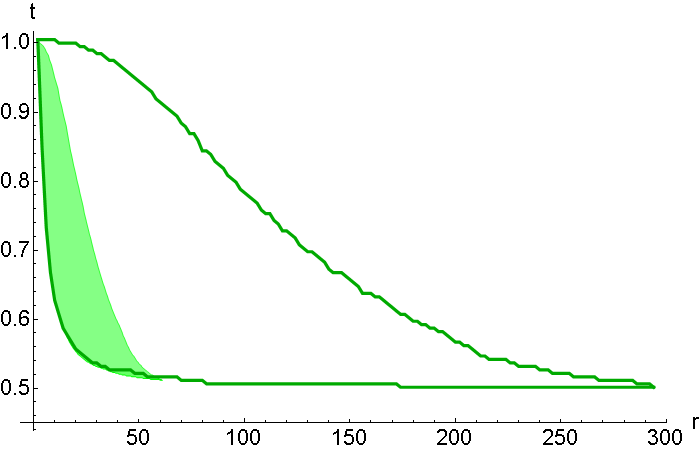}
&
\includegraphics[width=8cm]{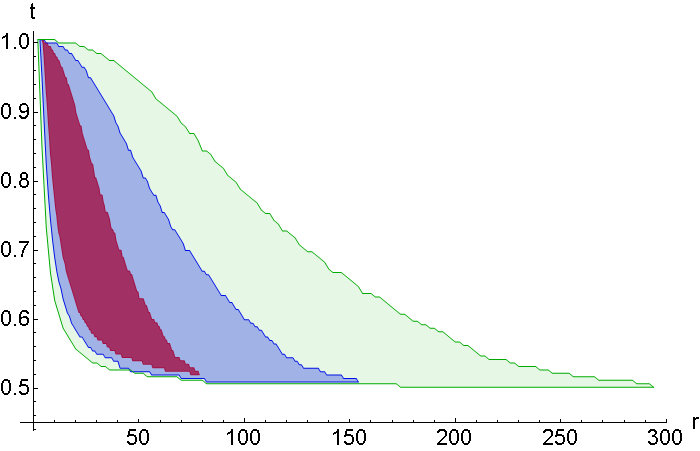}
\\
(a)&(b)
\end{tabular}
\caption{(a) For $|F/X| = 100 \tev$ and $d = 0.04 = 4 \times \Lambda/\Lambda_{UV}$ in Scenario 1, $V_{eff}$ has a local minimum in area of the $r$-$t$ plane enclosed by the green curve. For Scenario 2 this area shrinks down to the shaded region due to the increased effect of the instanton term. (b) Areas of the $r$-$t$ plane where $V_{eff}$ has a local minimum for $d = 0.04, 0.02, 0.01$ (green/light, blue/medium, red/dark) in Scenario 1. $r_{max} \propto d^{5/6}$, so decreasing $d$ from $0.04$ to $0.01$ decreases the area where there is a minimum. These areas do not depend significantly on $h$.
}
\label{f.tplot}
\end{center}
\end{figure}

We can explain this behavior of $r_{max}$ analytically. For fixed other parameters, $r_{max}$ is approximately the value of $r$ for which the scale of the gradient of $V_\mathrm{CW}$ near the minimum becomes smaller than the scale of the gradient of $V_{tree}$ (\eref{VtreeInstanton}). Therefore, we can roughly estimate $r_{max}$ by equating the gradient of the leading log approximation to $V_\mathrm{CW}$ (\eref{Vcwleadinglog}) to the gradient of $V_{tree}$ for $M_{sing} \sim \sqrt{h/d} f$ and $t \sim 0.5$. This yields
\begin{equation}
\label{e.rmax}
r_{max} \sim d^{5/6} \left( \frac{\Lambda}{|F/X|}\right)^{1/3}
\end{equation}
and explains the approximate linear dependence of $r_{max}$ on $d$ observed numerically. For Scenarios 1 and 2 this gives $r_{max} \sim 10^2$ and $\sim 10^1$, depending on the exact value of $d$. This agrees with our numerical results to $\sim 30 \%$.

\begin{figure}
\begin{center}
\begin{tabular}{cc}
$\tilde d = 1$&
$\tilde d = 4$\\
\includegraphics[width=8cm]{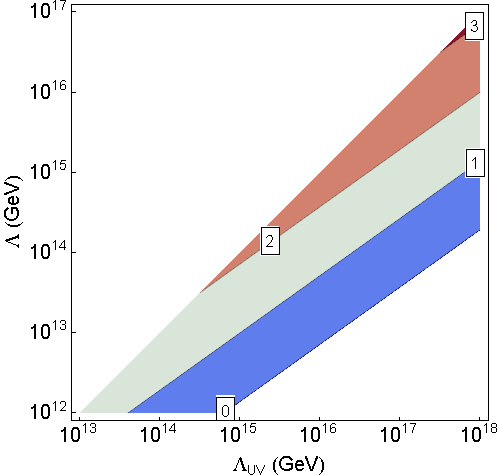}
&
\includegraphics[width=8cm]{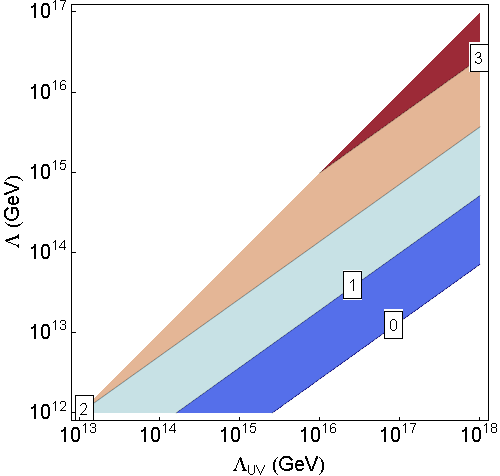}
\\
(a)&(b)
\end{tabular}
\caption{Estimate of $\log_{10} r_{max}$ for two possible values of $\tilde d = \frac{d}{\Lambda/\Lambda_{UV}}$. The upper and lower regions are excluded to satisfy $\Lambda < \Lambda_{UV}/100$ and $r > 1$ respectively. This demonstrates that the allowed range for $r$ shrinks to nothing for $\Lambda/\Lambda_{UV} \ll 1/1000$, making large $\Lambda$ heavily favored in our model.}
\label{f.rmaxplot}
\end{center}
\end{figure}

In Figure \ref{f.rmaxplot} we illustrate the range of allowed $r$-values by plotting the approximate $r_{max}$ from \eref{rmax} as a function of $\Lambda$ and $\Lambda_{UV}$. Since $r > 1$ is required for singlet VEVs, the shrinking of $r_{max}$ with decreasing $\Lambda$ effectively defines a minimum allowed value of $\Lambda/\Lambda_{UV}$, and for $\Lambda \lsim \Lambda_{UV}/1000$ it becomes very difficult to find a metastable uplifted vacuum because the allowed range of r shrinks to nothing. This means that $\Lambda$ as large as possible is favored in our model, and  justifies considering only our two scenarios with $\Lambda/\Lambda_{UV} \sim 1/100$.

Finally, we can also use these ideas to get a rough estimate of the pseudomodulus mass scale. Simply differentiating \eref{Vcwleadinglog} and setting $t \sim 0.5$ yields \eref{DDVcwestimate}.

%comment on rmax
% show rmax plot to show that these two scenarios are the only viable ones.
% note how we demonstrate effect of instanton term by artificially fixing F/X = 100 TeV exactly and numerically exploring when we have a vacuum.

\subsection{Validity of 1-loop calculation}
\label{ss.validityof1loop}
The smallness of  $d \sim 0.01$ compared to  $h \sim 1$ and $g$ (depending on the size of $r$) might cause us to suspect that all these results would be invalidated by 2-loop corrections. Fortunately, this naive expectation is not realized due to the nature of contributions to the effective potential. The leading-log approximation to the 1-loop potential \eref{Vcwleadinglog} is a very good approximation for the \emph{complete} contributions from messengers (loops involving the $h$-coupling) and singlets with $D$-term masses (involving the $g$-coupling), as well as the \emph{logarithmic} contributions from singlets with $F$-term masses. The only components not included are the small-$x$ contributions from singlets with $F$-term masses, and those are the contributions with non-trivial features required to generate the minimum.

The tuning can be understood as canceling the smooth logarithmic contributions to the effective potential to high enough precision so that the minimum created by the contributions from singlets with $F$-term masses survives.
Since $d$ is so small, this local minimum is pushed out to rather large field values $M_{sing} \sim \sqrt{h/d} \ f$ where the leading log approximation for the `uninteresting' contributions is excellent. This makes the two-loop corrections involving two $h$ and $g$ couplings (messengers and singlets with $D$-term masses, respectively) very smooth as well, meaning they do not introduce any gross new features to the effective potential. Therefore they just generate a smooth correction to \eref{Vcwleadinglog}, which can be compensated for by slightly adjusting the gauge coupling $g$ (or the ratio $\mprime/f$) and should not significantly affect the existence of local minima or the severity of tuning (though \eref{mprimetuning} might have to be slightly adjusted). Therefore the important features of our analysis are valid.

\subsection{Lifetime Constraints on Uplifted Vacuum Stabilization}
\label{ss.lifetimecalculation}
% drastically reduce size of this section, basically show that bounce action constraints are always OK due to small d and large Lambda.
We now check that the uplifted vacuum is stable enough to have not decayed in the lifetime of the universe. For each decay path across the potential landscape we estimate the Bounce Action $B$  which exponentially suppresses the decay width \cite{FateOfTheFalseVacuum}. We require $B \gsim 10^3$ \cite{requiredsizeofB}. For rough estimates of the bounce action we approximate the potential along the decay path as a triangular barrier, which yields very simple analytical expressions for $B$ \cite{TrianglePotential}.

There are two decay paths that are only forbidden by loop-sized effects. As illustrated in \fref{vacuumstructure}, $M_{sing}$ can either tunnel towards the origin, in which case the messengers become tachyonic and the fields roll towards the ISS vacuum, or it can tunnel away from the origin and roll towards the SUSY-minimum.

To estimate the bounce action for decay to the ISS vacuum along the pseudoflat direction we take limit where the height of the potential barrier and the distance from the edge of the barrier to the ISS vacuum goes to zero. This \emph{underestimates} $B$ and gives
\begin{equation}
B_\mathrm{ISS} > 2 \pi^2 \frac{N_f-1}{N_f} \frac{r^4 (2 r^2-1)^2}{(d/b^2)^2} \sim \underbrace{\frac{8 \pi^2}{5}}_{\sim 15} \underbrace{\left(\frac{\Lambda_{UV}}{\Lambda}\right)^2}_{> 10^4} b^4 \underbrace{r^4 \left( 2 r^2 - 1\right)^2}_{>1} \gg 10^3
\end{equation}
Turning to the bounce action for decay to the SUSY vacuum along the pseudoflat direction we again take the height of the potential barrier to zero and neglect several unknown or parametrically smaller contributions to the length of the decay path. Using $\Delta V^0$ from \eref{ISSGKKenergydifference} as the depth of the potential well on the other side of the barrier we obtain (neglecting $O(1)$ factors)
\begin{equation}
B_\mathrm{SUSY} > \frac{32 \pi^2}{3}  \sqrt{\frac{\Lambda}{f}} \frac{1}{d^{3/2}} \gg 10^3
\end{equation}
Both decays are sufficiently suppressed.

%%%%%%%%%%%%%%%%%%%%%%%%%%%%%%%%%%%%%%%%%%%%%%%%%%%%%%%%%%%%%%%%%%%%%%
%%%%%%%%%%%%%%%%%%%%%%%%%%%%%%%%%%%%%%%%%%%%%%%%%%%%%%%%%%%%%%%%%%%%%%%
\section{Conclusions}
\label{s.conclusions} \setcounter{equation}{0} \setcounter{footnote}{0}
%%%%%%%%%%%%%%%%%%%%%%%%%%%%%%%%%%%%%%%%%%%%%%%%%%%%%%%%%%%%%%%%%%%%%%
%%%%%%%%%%%%%%%%%%%%%%%%%%%%%%%%%%%%%%%%%%%%%%%%%%%%%%%%%%%%%%%%%%%%%
The ISS framework \cite{ISS} is an extremely appealing model building arena for exploring non-perturbative meta-stable SUSY-breaking. However, previous ISS-based  models of Direct Gauge Mediation are plagued by several problems, both aesthetic and phenomenological, which include small gaugino masses (exacerbating the little hierarchy problem), Landau Poles and non-renormalizable operators with somewhat contrived flavor contractions. Since the issue of small gaugino masses has been understood to be related to the vacuum structure of the theory \cite{GKK}, one model-building challenge is the formulation of plausible uplifted ISS models.

We first outlined some simple but general model-building guidelines for stabilizing uplifted ISS models, which lead us to conclude that meson-deformations are required (or at least heavily favored) to stabilize the adjoint component of the magnetic meson in the hidden sector. However, the singlet can be stabilized by a variety of mechanisms, which makes it possible that an uplifted hidden sector with minimal flavor group might be viable.

This lead us to propose Singlet Stabilized Minimal Gauge Mediation as a simple ISS-based model of Direct Gauge Mediation which avoids both light gauginos and Landau Poles. The hidden sector has trivial magnetic gauge group and minimal unbroken $SU(5)$ flavor group, while the uplifted vacuum is stabilized by a singlet sector with its own $U(1)$ gauge symmetry, generating a nonzero VEV for the singlet meson via the inverted hierarchy mechanism.

The stabilization mechanism used in our model necessitates adjusting parameters to a precision of $\sim (\Lambda/\Lambda_{UV})^2 \sim 10^{-4}$, a common problem with quantum-stabilized models. While this tuning can be reduced by being less conservative about the separation of scales, one might question the advantage of this tuning compared to the tuning in the MSSM higgs-sector associated with a split-SUSY spectrum. Apart from the fact that a split-SUSY spectrum might not be experimentally observed, the key is that a split-SUSY spectrum \emph{cannot be avoided} in most models of Direct Gauge Mediation that are in the ground state, in particular standard ISS\footnote{One might have an independent suppression mechanism for the sfermion masses, see for example \cite{DirectGauginoMediation} }. This paper shows that it is possible to stabilize an \emph{uplifted} ISS model with \emph{very small flavor group}, a necessary condition for avoiding Landau Poles of the SM gauge couplings, and while the current stabilization mechanism requires said tuning it seems plausible that an alternative mechanism with generically stabilized uplifted vacua exists. That makes our stabilization-tuning preferable to the `unavoidable' higgs-sector tuning from a split-SUSY spectrum.

\subsection*{Acknowledgements}
We are extremely grateful to Csaba Csaki, Zohar Komargodski, Maxim Perelstein and Liam McAllister for valuable insights and comments on the manuscript. We would also like to thank Markus Luty, John Terning, Jesse Thaler,  David Shih, Rouvan Essig, Nathan Seiberg, Andrey Katz and Flip Tanedo for helpful discussion. The work of D.C. and Y.T. was supported in part by the National Science Foundation under grant PHY-0355005. D.C. was also supported by the John and David Boochever Prize Fellowship in Fundamental Theoretical Physics 2010-11. Y.T. is also supported by a Fermilab Fellowship in Theoretical Physics.
Fermilab is operated by Fermi Research Alliance, LLC, under Contract No.~DE-AC02-07CH11359 with the United States Department of Energy.

\end{document}